\documentclass[preprint,12pt,nonatbib]{elsarticle}

\usepackage[T1]{fontenc}
\usepackage[utf8]{inputenc}
\usepackage[english]{babel}
\usepackage{amsmath,amssymb,amsfonts,mathtools,bm,mathrsfs}
\usepackage{graphicx}
\usepackage{subcaption}
\usepackage{booktabs}
\usepackage{multirow}
\usepackage{siunitx}
\usepackage{algorithm}
\usepackage{algorithmicx}
\usepackage{algpseudocode}
\usepackage{xcolor}
\usepackage{microtype}
\usepackage{lineno}
\usepackage{hyperref}

\definecolor{custom-blue}{RGB}{3,69,173}
\hypersetup{colorlinks=true,urlcolor=blue,citecolor=custom-blue,linkcolor=custom-blue}

\begin{document}
\begin{frontmatter}
\journal{Engineering Structures}
\title{Generative Learning on Manifolds for Probabilistic Seismic Performance Assessment and Structural Identifiability}
\author[jhu_cse,jhu_ams]{Dimitris G. Giovanis\corref{cor1}}
\ead{dgiovan1@jhu.edu}
\author[jhu_cbe,jhu_ams]{Ioannis G. Kevrekidis}
\ead{yannisk@jhu.edu}
\author[usc_cee,usc_ame]{Roger G. Ghanem}
\ead{ghanem@usc.edu}
\cortext[cor1]{Corresponding author}
\affiliation[jhu_cse]{organization={Dept. of Civil \& Systems Engineering}, organization={Johns Hopkins University}}
\affiliation[jhu_cbe]{organization={Dept. of Chemical and Biomolecular Engineering}, organization={Johns Hopkins University}}
\affiliation[jhu_ams]{organization={Dept. of Applied Mathematics and Statistics}, organization={Johns Hopkins University}}
\affiliation[usc_cee]{organization={Dept. of Civil and Environmental Engineering}, organization={University of Southern California}}
\affiliation[usc_ame]{organization={Dept. of Aerospace and Mechanical Engineering}, organization={University of Southern California}}

\begin{abstract}
Seismic performance assessment relies on computationally expensive nonlinear simulations, limiting uncertainty quantification and inference of structural properties from observed responses. We develop a manifold-based generative framework that learns the joint distribution of uncertain structural parameters and seismic-response trajectories from limited high-fidelity simulations. Combining Diffusion Maps,  score-based Diffusion Modeling, and Geometric Harmonics, the framework enables both forward probabilistic prediction and inverse inference. Applied to a nine-story steel moment-resisting frame, the results reveal which structural properties are identifiable from seismic performance and how uncertainty in these properties influences response. In particular, the learned representation captures post-capping degradation characteristics more effectively than individual strength parameters. Inverse inference identifies distinct structural configurations that produce similar seismic performance, with representative solutions independently validated using nonlinear simulations. These findings demonstrate how generative learning can support uncertainty-aware seismic assessment while revealing structural non-identifiability and alternative mechanisms underlying observed performance.
\end{abstract}

\end{frontmatter}

\section{Introduction}
\label{S:Background}

\noindent
Natural hazard risk reduction increasingly depends on computational models that connect hazard intensity to the performance of exposed infrastructure. In earthquake engineering, performance-based assessment combines seismic hazard characterization, nonlinear structural analysis, and consequence evaluation to quantify how buildings and infrastructure respond across a range of intensity levels and uncertainty sources \cite{stojadinovici2018case,li2019probabilistic}. A central computational challenge is that nonlinear response-history analysis (NLRHA) must be repeated over many ground-motion records, intensity levels, and uncertain structural models. This becomes particularly expensive in Incremental Dynamic Analysis (IDA), where each ground motion is repeatedly scaled until the response evolves from the elastic regime toward severe nonlinearity and collapse.

Artificial intelligence and machine learning methods offer an opportunity to reduce this computational burden, but natural hazard applications require more than point prediction. Reliable models must preserve uncertainty, remain meaningful when high-fidelity data are scarce, and respect the structure imposed by the underlying physics. In performance-based earthquake engineering, both epistemic uncertainty in structural modeling and record-to-record variability in ground motion can strongly influence response, especially in the nonlinear and near-collapse regimes \cite{vamvatsikos2010incremental}. Existing acceleration strategies include advanced Monte Carlo methods \cite{vamvatsikos2014seismic,patsialis2021multi}, reduced-order modeling \cite{patsialis2020reduced}, fragility-estimation procedures \cite{jalayer2020seismic}, semi-analytical stochastic methods \cite{chen2010stochastic,dos2016incremental}, and surrogate models \cite{gidaris2015kriging,zou2020surrogate,abbiati2021seismic,kyprioti2021kriging,giovanis2016epistemic,pang2014artificial,bakalis2017seismic,zhu2023seismic,yi2024stochastic,giovanis2025accelerating}. Most surrogate approaches, however, are formulated primarily as deterministic or conditionally averaged input--output maps. Such models can be highly effective for prediction but do not directly provide a generative representation of the coupled distribution of uncertain structural properties and fixed-dimensional seismic-response representations.

Generative models provide a complementary perspective by learning probability distributions rather than only deterministic mappings. Score-based diffusion models are particularly attractive because they represent complex target distributions through a diffusion process and corresponding reverse dynamics \cite{song2020score,song2019generative,song2020improved,ho2020denoising}. Their direct application can nevertheless be difficult when high-dimensional engineering data are concentrated near a lower-dimensional manifold and only a limited number of high-fidelity samples are available. Probabilistic Learning on Manifolds and related nonlinear manifold-learning methods address this setting by identifying intrinsic coordinates and performing probabilistic learning in the reduced representation \cite{soize2016data,soize2022probabilistic,coifman2006diffusion,giovanis2025enabling}. Recent work has combined score-based generative learning with Diffusion Maps and Double Diffusion Maps to sample distributions supported near learned manifolds and reconstruct ambient variables through Geometric Harmonics \cite{giovanis2025generative,crabtree2025generative,evangelou2023double,coifman2006geometric}.

Here we adapt this manifold-based generative framework to probabilistic seismic performance assessment. IDA is especially well suited to this setting because each uncertain structural realization produces a response trajectory rather than a single scalar output. Instead of approximating selected response quantities independently, we learn the \emph{joint probability distribution} of uncertain structural parameters and the corresponding representative IDA response. The resulting representation supports both unconditional generation of structural parameter--response pairs and forward conditional inference of response distributions for prescribed structural properties. The principal contributions of this work are:
\begin{enumerate}

\item We formulate probabilistic seismic performance assessment through the joint distribution of uncertain structural parameters and representative IDA responses, enabling uncertainty-aware inference beyond conventional deterministic input--output surrogate models.

\item We develop a hybrid manifold-based generative framework that retains structural parameters explicitly while representing seismic-response trajectories through non-harmonic (essentially independent) Diffusion Maps coordinates. Combined with score-based diffusion modeling, this enables probabilistic learning from limited high-fidelity simulations without additional nonlinear analyses for generation.

\item We investigate the capability of accurately inferring the structural parameter
values of the model from observations of its behavior (``data'').


\item We demonstrate forward and inverse probabilistic inference using the learned input-output joint distribution. Forward conditioning characterizes the influence of uncertain structural properties on seismic performance, while inverse conditioning reveals distinct structural configurations compatible with the same observed response. 

\end{enumerate}

The proposed framework is demonstrated using a nine-story steel moment-resisting frame \cite{foutch2002modeling} with uncertain nonlinear structural properties, previously investigated in the context of incremental dynamic analysis (IDA) and uncertainty quantification \cite{giovanis2016epistemic,giovanis2025accelerating}. Unlike our previous studies, which primarily focused on forward surrogate prediction and dimensionality reduction \cite{giovanis2016epistemic,giovanis2025accelerating}, the present work aims to learn the \textit{joint parameter--response distribution}, enabling generative sampling, conditional probabilistic inference, and the investigation of parameter identifiability. In this context, identifiability concerns whether distinct combinations of structural parameters produce distinguishable structural responses. Specifically, if the forward mapping from model parameters to structural responses is injective, the parameters can, in principle, be uniquely recovered from the observed responses, and the model is said to be structurally identifiable. Conversely, if multiple parameter configurations yield identical responses, the inverse problem is nonunique, indicating structural unidentifiability. From an analytical perspective, local identifiability can be investigated by examining the local injectivity of the forward mapping, typically through the rank of its Jacobian with respect to the uncertain parameters. The proposed probabilistic framework complements this perspective by characterizing the conditional distribution of structural parameters given observed responses, thereby capturing parameter ambiguity, uncertainty, and potentially multiple plausible parameter configurations.


The remainder of the paper introduces the score-estimation and manifold-learning components, formulates the joint generative problem, presents the earthquake-engineering application and validation results, and concludes with implications, limitations, and directions for uncertainty-aware AI in seismic performance assessment.

\section{Methods}

\subsection{Score-Based Generative Modeling}
\label{sec:SGM}

\noindent
Score-based generative models define a forward stochastic differential equation
(SDE) that progressively transforms a target data distribution into a tractable
noise distribution and then use the corresponding reverse dynamics to generate
samples from the target law \cite{song2020score}. A general forward SDE can be
written as
\begin{equation}
    \mathrm{d}\mathbf{x}
    =
    f(\mathbf{x},t)\,\mathrm{d}t
    +
    g(t)\,\mathrm{d}\mathbf{B}_t,
    \label{eq:sde}
\end{equation}
where $f(\mathbf{x},t)$ and $g(t)$ are the drift and diffusion terms and
$\mathbf{B}_t$ is Brownian motion. The associated reverse-time SDE depends on
the score $\nabla_{\mathbf{x}}\log p_t(\mathbf{x})$,
\begin{equation}
    \mathrm{d}\mathbf{x}
    =
    \left[
    f(\mathbf{x},t)
    -
    g(t)^2\nabla_{\mathbf{x}}\log p_t(\mathbf{x})
    \right]\mathrm{d}t
    +
    g(t)\,\mathrm{d}\bar{\mathbf{B}}_t.
    \label{eq:reversesde}
\end{equation}
Conventional diffusion models approximate the score with a neural network
trained by score matching or denoising score matching
\cite{vincent2011connection,songdiffmodels}.

In this work, we instead use the Monte Carlo score-estimation strategy of
Liu et al.~\cite{liu2024diffusion}. For a forward process with Gaussian
transition density
\begin{equation}
p(\mathbf{x}_t\mid\mathbf{x}_0)
=
\mathcal{N}
\left(
\alpha_t\mathbf{x}_0,
\beta_t^2\mathbf{I}_d
\right),
\end{equation}
the score can be estimated directly from the available reference samples
without training a separate score network. Denoting the fixed empirical
reference set by $\{\mathbf{x}^n\}_{n=1}^{N}$, the estimator is
\begin{equation}
\widehat{\mathbf{s}}(\mathbf{x}_t,t)
=
\sum_{n=1}^{N}
\frac{\alpha_t\mathbf{x}^n-\mathbf{x}_t}{\beta_t^2}
\,\overline{w}_t(\mathbf{x}_t,\mathbf{x}^n),
\qquad
\overline{w}_t
=
\frac{
p(\mathbf{x}_t\mid\mathbf{x}^n)
}{
\sum_{m=1}^{N}
p(\mathbf{x}_t\mid\mathbf{x}^m)
}.
\label{eq:mc_score}
\end{equation}

For the generative state $\mathbf{z}$ used in the present implementation,
the Gaussian forward transition is parameterized as
\begin{equation}
p(\mathbf{z}_t\mid\mathbf{z}_0)
=
\mathcal{N}
\left(
\alpha_t\mathbf{z}_0,
\beta_t^2\mathbf{I}
\right),
\end{equation}
with
\begin{equation}
\alpha_t = 1-\gamma t,
\qquad
\beta_t^2 = (1-\gamma)+\gamma t,
\qquad
t\in[0,1],
\label{eq:alpha_beta}
\end{equation}
where $\gamma=0.999$. Thus,
$\alpha_0=1$, $\beta_0^2=10^{-3}$,
$\alpha_1=10^{-3}$, and $\beta_1^2=1$.
The nonzero value of $\beta_0^2$ provides a small variance floor that
regularizes the empirical score near the data endpoint $t=0$. The corresponding linear forward drift coefficient is
\begin{equation}
b(t)
=
\frac{\dot{\alpha}_t}{\alpha_t}
=
-\frac{\gamma}{1-\gamma t},
\end{equation}
and the diffusion variance satisfies
\begin{equation}
\sigma^2(t)
=
\frac{\mathrm{d}\beta_t^2}{\mathrm{d}t}
-
2b(t)\beta_t^2
=
\gamma-2b(t)\beta_t^2.
\label{eq:sigma}
\end{equation}
The associated probability-flow ODE is
\begin{equation}
\frac{\mathrm{d}\mathbf{z}_t}{\mathrm{d}t}
=
b(t)\mathbf{z}_t
-
\frac{1}{2}\sigma^2(t)
\nabla_{\mathbf{z}}
\log p_t(\mathbf{z}_t).
\label{eq:probability_flow}
\end{equation}

Generation starts at $T=1$ from
$\mathbf{z}_1\sim\mathcal{N}(\mathbf{0},\mathbf{I})$,
which approximates the terminal marginal since $\alpha_1=10^{-3}$, and
Eq.~\eqref{eq:probability_flow} is integrated backward toward $t=0$.
A uniform time grid
\begin{equation}
t_k = 1-k\Delta t,
\qquad
\Delta t=\frac{1}{N_t},
\qquad
k=0,\ldots,N_t,
\end{equation}
is used with $N_t=1000$. With explicit Euler integration,
\begin{equation}
\mathbf{z}_{k+1}
=
\mathbf{z}_k
-
\Delta t
\left[
b(t_k)\mathbf{z}_k
-
\frac{1}{2}\sigma^2(t_k)
\widehat{\mathbf{s}}(\mathbf{z}_k,t_k)
\right].
\label{eq:ode_discrete}
\end{equation}
The drift and score are evaluated for $k=0,\ldots,N_t-1$; therefore, the
final update reaches $t=0$ without evaluating the score at the endpoint
itself.

\subsection{Manifold-Based Generative Learning of Seismic Response}
\label{sec:method}

\noindent
We couple physics-based IDA with manifold-based generative learning to estimate a low-dimensional representation of the dominant coupled variability between uncertain structural properties and aggregated seismic-response trajectories. Let
\begin{equation}
\mathbf{x}_s \in \mathbb{R}^{n_s},
\qquad
\mathbf{x}_s \sim f_{\mathbf{x}_s}(\cdot),
\label{eq:structural_parameters}
\end{equation}
denote the uncertain structural-parameter vector, and let
\begin{equation}
\mathbf{x}_g^{(r)} \in \mathbb{R}^{n_g},
\qquad
r=1,\ldots,N_g,
\label{eq:ground_motion_parameters}
\end{equation}
denote the $r$th ground-motion realization. The structural model evaluated within IDA is written as
\begin{equation}
\mathcal{G}:
\left(
\mathbf{x}_s,
\mathbf{x}_g^{(r)},
\lambda
\right)
\longmapsto
\left(
D^{(r)}(\lambda),
S_a^{(r)}(\lambda)
\right),
\label{eq:structural_model}
\end{equation}
where $\lambda$ denotes the record scaling level, $D^{(r)}(\lambda)$ is the corresponding maximum interstory drift, and $S_a^{(r)}(\lambda)$ is the associated 5\%-damped first-mode spectral acceleration. For each structural realization, $N_g=30$ single-record IDA curves are available. To obtain a fixed-dimensional representation, each record-level IDA is first converted to a single-valued first-passage relation \(S_a^{(r)}(D)\) and interpolated onto a common drift grid. The representative response for each structural realization is then defined as the pointwise median of the 30 interpolated record-level responses. For records terminating before the end of the common grid, the terminal spectral-acceleration value is retained over the remaining drift coordinates.
The 30 record-level responses are then aggregated pointwise on the common drift grid.  Let $\mathbf{D}^{\mathrm{com}}
=
[
D_1^{\mathrm{com}},\ldots,D_{18}^{\mathrm{com}}
]$ denote the common drift grid, consisting of 18 prescribed drift levels between \(0\) and \(0.20\). At each drift coordinate $D_j^{\mathrm{com}}$, we compute
\begin{equation}
\widetilde{S}_{a,j}
=
\operatorname{median}_{r=1,\ldots,N_g}
S_a^{(r)}\!\left(D_j^{\mathrm{com}}\right),
\qquad
j=1,\ldots,18,
\label{eq:aggregated_response_coordinates}
\end{equation}
with $N_g=30$.The representative response is then defined pointwise as
\begin{equation}
\widetilde{\mathbf{S}}_a
=
\left[
\widetilde{S}_{a,1},\ldots,\widetilde{S}_{a,18}
\right]
\in\mathbb{R}^{18}.
\label{eq:median_acceleration_vector}
\end{equation}
The representative response is therefore a monotone median IDA curve evaluated on the common drift grid,  written as $\mathbf{x}_r
=
\widetilde{\mathbf{S}}_a
\in\mathbb{R}^{18}$. The learning vector in physical coordinates is then defined as
\begin{equation}
\mathbf{x}
=
\left(
\mathbf{x}_s,
\mathbf{x}_r
\right)
=
\left(
\mathbf{x}_s,
\widetilde{\mathbf{S}}_a
\right)
\in
\mathbb{R}^{n_s+18}.
\label{eq:augmented_vector}
\end{equation}
For the present application, $n_s=6$, so that $\mathbf{x}\in\mathbb{R}^{24}.$
Ground-motion record-to-record variability enters the analysis through the 30 single-record IDA trajectories used to construct $\mathbf{x}_r$; the ground-motion variables and individual record-level responses are not explicit random variables in the learned generative representation.

Building on our previous manifold-constrained generative framework \cite{giovanis2025generative}, the present implementation combines Diffusion Maps~\cite{coifman2006geometric} with training-free Monte Carlo score estimation~\cite{liu2024diffusion}. In contrast to applying dimensionality reduction to the complete structural parameter--response vector, Diffusion Maps are applied only to the 18-dimensional representative response. The six structural parameters remain explicit coordinates throughout the generative procedure. This avoids reconstructing independently sampled structural variables through the response-manifold lifting map while retaining a low-dimensional representation of the nonlinear IDA response. Starting from the training response vectors
\begin{equation}
\mathbf{X}_r
=
\left\{
\mathbf{x}_{r,i}
\right\}_{i=1}^{N_{\mathrm{tr}}}
\subset
\mathbb{R}^{18},
\end{equation}
we apply Diffusion Maps \cite{coifman2006diffusion} to obtain a response latent representation
\begin{equation}
\boldsymbol{\Phi}_r
=
\left\{
\boldsymbol{\phi}_{r,i}
\right\}_{i=1}^{N_{\mathrm{tr}}}
\subset
\mathbb{R}^{d_r}.
\label{eq:latent_representation}
\end{equation}
Fifteen diffusion eigenpairs are evaluated, and local regression residuals are used to identify non-harmonic coordinates. Six response coordinates are retained. The objective is not strictly dimensionality reduction, but identification of a compact set of non-redundant coordinates that preserves the intrinsic response geometry.  
The generative state is then defined as
\begin{equation}
\mathbf{z}_i
=
\left(
\mathbf{x}_{s,i}^{(\mathrm{std})},
\boldsymbol{\phi}_{r,i}
\right)
\in
\mathbb{R}^{6+6}
=
\mathbb{R}^{12}.
\label{eq:generative_state}
\end{equation}
Although the response representation is reduced only from 18 to 6 coordinates, the effect on the joint generative problem is more substantial: the physical parameter--response space is reduced from 24 to 12 dimensions. This reduction is particularly important in the present small-sample regime, where only 151 training realizations are available. Reducing the effective dimension decreases the difficulty of estimating the empirical score and the underlying joint distribution from a limited number of samples, while the held-out reconstruction results indicate that the dominant response information is retained.

\begin{figure}[!htb]
     \centering
     \includegraphics[width=1.0\linewidth]{ 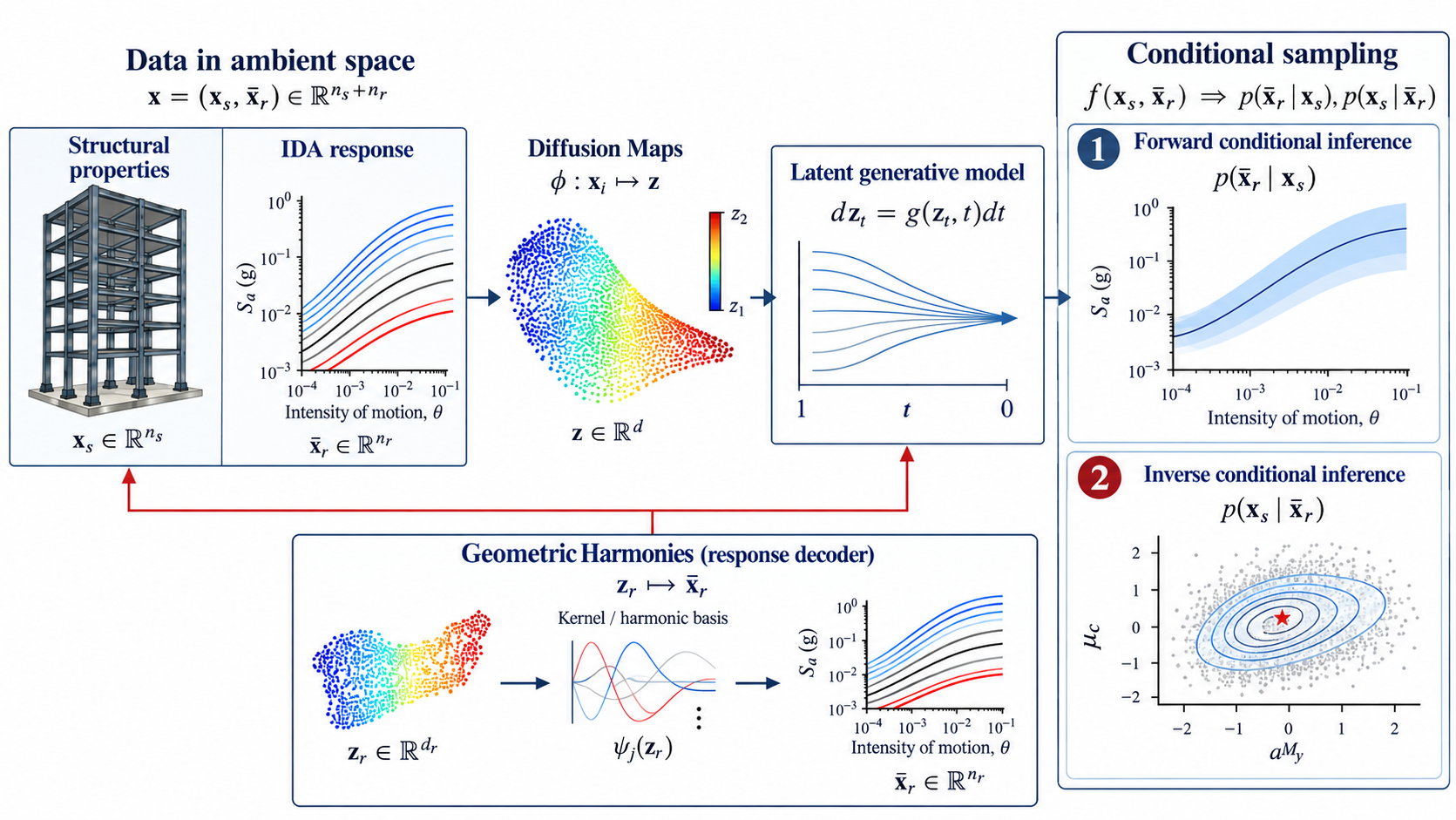}
\caption{Proposed manifold-based generative framework for learning the joint distribution of structural properties and seismic response and enabling conditional sampling.}
     \label{fig:method}
 \end{figure}

In the generative phase, samples
$\mathbf{y}\sim\mathcal{N}(\mathbf{0},\mathbf{I}_{12})$
are drawn and transported directly through the probability-flow ODE using the Monte Carlo score estimate. The generated structural coordinates are transformed directly back to physical units. Only the generated response latent coordinates are lifted to the 18-dimensional response space using Geometric Harmonics (GH) (see \ref{sec:GH}). The resulting samples
\begin{equation}
\left\{
\widetilde{\mathbf{x}}_i
\right\}_{i=1}^{N^\star}
=
\left\{
\left(
\widetilde{\mathbf{x}}_{s,i},
\widetilde{\mathbf{x}}_{r,i}
\right)
\right\}_{i=1}^{N^\star}
\sim
f_{\mathbf{x}_s,\mathbf{x}_r}(\cdot)
\label{eq:generated_ambient_samples}
\end{equation}
provide an expanded ensemble of structural parameter--response realizations. A schematic illustration of the approach is provided in Fig.~\ref{fig:method}.

\subsection{Earthquake Application: Nine-Story Steel Moment-Resisting Frame}
\label{sec:LA9}

\noindent
We consider a nine-story steel moment-resisting frame with a single-story basement, hereafter referred to as LA9, designed for Los Angeles in accordance with the 1997 NEHRP provisions \cite{foutch2002modeling}. The structure is represented by a centerline model developed in OpenSees, incorporating nonlinear beam--column connections that allow plastic hinges to form at the beam ends, while the columns are assumed to remain elastic. While uncertainties in material yield strengths may cause columns to yield before their connecting beams, previous analyses have shown that this effect is relatively minor for the LA9 structure, particularly when beam and column steel strengths are strongly correlated \cite{vamvatsikos2010incremental}. Moreover, the authors in \cite{fragiadakis2006evaluation} demonstrated that plastic-hinge yield strength has a greater influence on displacement-related structural responses than variations in story mass and stiffness, highlighting its importance in structural uncertainty quantification. $P$--$\Delta$ effects and the internal gravity frames are included in the
model. The fundamental period of the reference structure is
$T_1=2.35$~s, with the first mode accounting for approximately $84\%$ of the
total mass. The structure is therefore predominantly first-mode dominated,
while retaining sensitivity to higher-mode effects. The present study focuses
on uncertainty in the beam-hinge properties governing drift-sensitive
structural response.

\begin{figure}[htbp]
    \centering
    \begin{subfigure}[b]{0.49\textwidth}
        \centering
        \includegraphics[width=\textwidth]{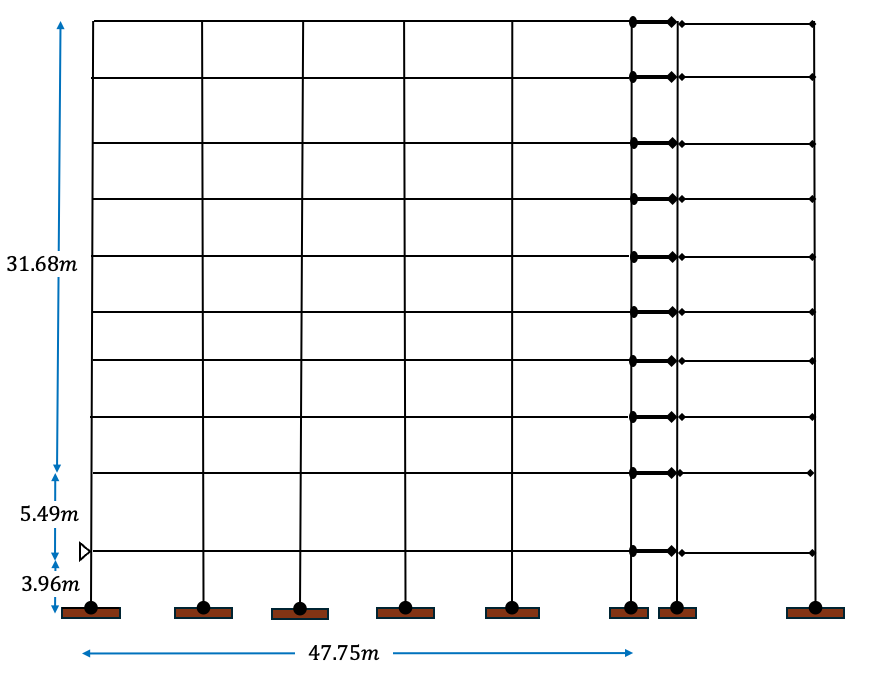} 
        \caption{}
        \label{fig:LA9a}
    \end{subfigure}
    \hfill
    \begin{subfigure}[b]{0.49\textwidth}
        \centering
        \includegraphics[width=\textwidth]{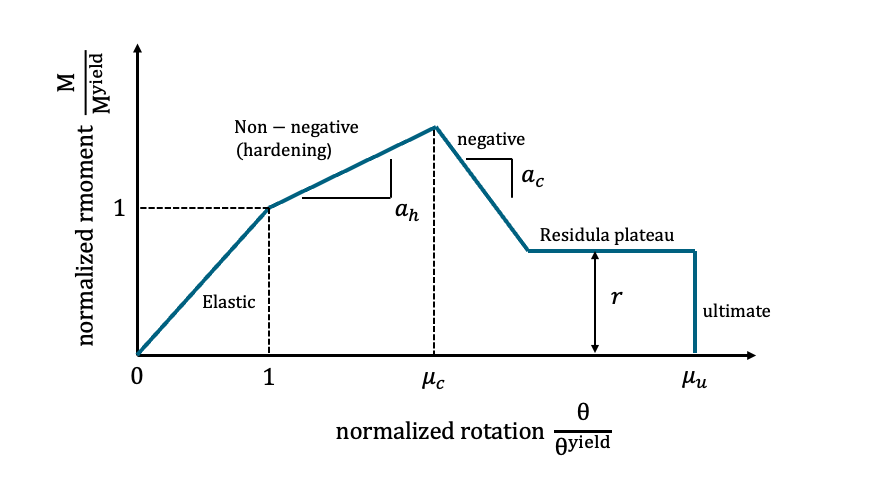} 
        \caption{}
        \label{fig:LA9b}
    \end{subfigure}
    
    \caption{(a) The LA9 steel moment-resisting frame and (b) The moment-rotation beam-hinge backbone (adopted from \cite{giovanis2025accelerating})}
    \label{fig:side-by-side}
\end{figure}

\subsubsection{Structural Properties}

\begin{figure}[!htb]
\centering
\includegraphics[width=0.9\textwidth]{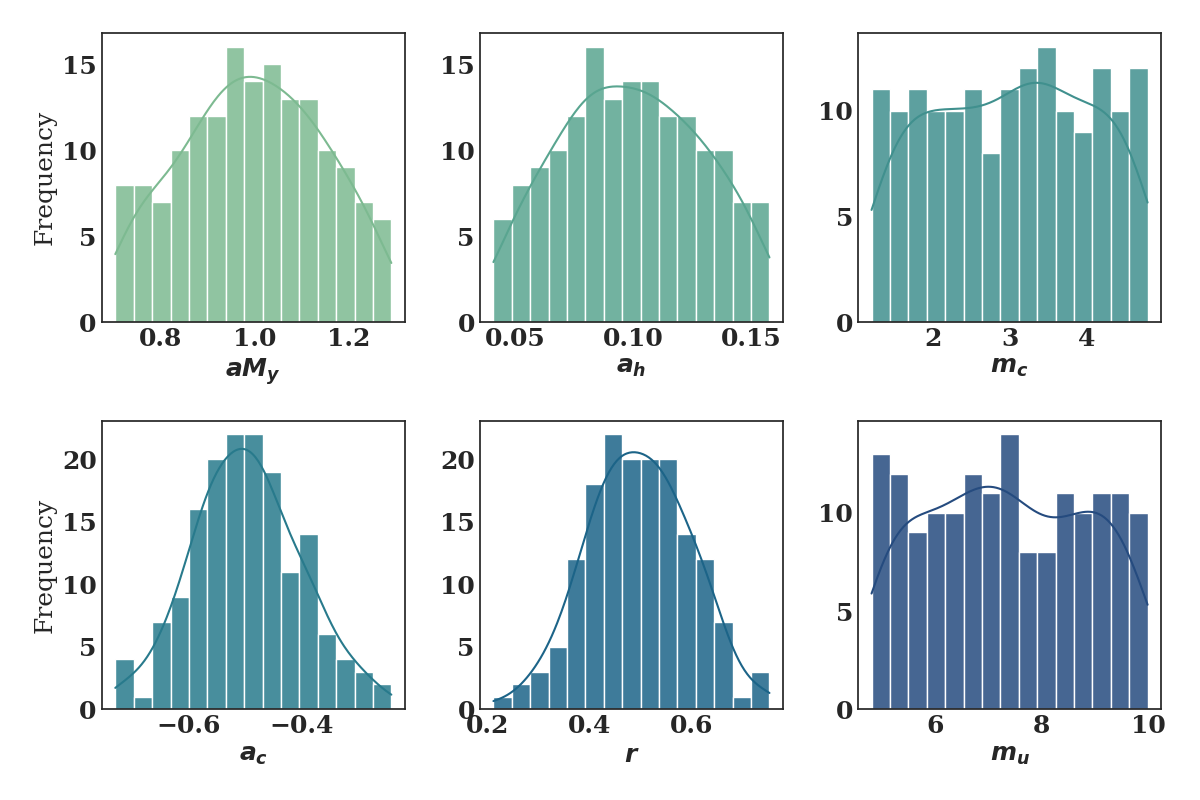}
\caption{Histograms of the 160 realizations of the structural-parameter vector $\mathbf{x}_s$ generated using Latin hypercube sampling \cite{iman1982distribution}.}
\label{fig:ida_samples}
\end{figure}

\noindent
The beam hinges\footnote{A beam hinge is a localized region of plastic deformation that allows the beam to rotate under excessive loading while maintaining load-carrying capacity, thereby promoting ductile behavior and preventing sudden failure.} were modeled as rotational springs with a quadrilinear moment-rotation backbone that exhibits symmetry for both positive and negative rotations.  The monotonic envelope of the hinge moment-rotation relationship exhibits hardening beyond a normalized yield moment $a_{M_y}$, with a non-negative slope $a_h$ extending up to a normalized rotation (or rotation ductility) $\mu_c$, where a negative stiffness segment begins. This drop, occurring at a slope $a_c$, is halted by a residual plateau at a normalized height $r$, which terminates abruptly at the ultimate rotational ductility $\mu_u$.  In this study, we consider six normalized parameters defining the moment--rotation backbone,
\begin{equation}
\mathbf{x}_s=[a_{M_y},a_h,\mu_c,a_c,r,\mu_u].
\end{equation}
The six parameter types are modeled as independent random variables following truncated Gaussian distributions to prevent non-physical values. Spatial variability among beam hinges is neglected; for each structural realization, the sampled parameter vector is assigned consistently to all corresponding beam hinges in the frame.
Therefore, $\mathbf{x}_s \in \mathbb{R}^6$. To generate the initial data, we employed Latin Hypercube Sampling (LHS) \cite{iman1982distribution} to  draw samples from $f_{\mathbf{x}_s}$.
Figure~\ref{fig:ida_samples} shows the 160 LHS realizations
drawn from the truncated-normal parameter distributions summarized in
Table~\ref{tab:structural_parameters}. The different histogram shapes reflect
the combined effects of the prescribed coefficients of variation, truncation
bounds, and finite LHS sampling.

\begin{table}[!htb]
\centering
\caption{Distribution properties of the uncertain structural parameters.}
\label{tab:structural_parameters}
\begin{tabular}{lccccc}
\hline
Parameter & Mean & COV (\%) & Lower bound & Upper bound & Distribution \\
\hline
$a_{M_y}$ & 1.0  & 20 & 0.70  & 1.30  & Truncated normal \\
$a_h$     & 0.10 & 40 & 0.04  & 0.16  & Truncated normal \\
$\mu_c$   & 3.0  & 40 & 1.20  & 4.80  & Truncated normal \\
$a_c$     & -0.50 & 40 & -0.80 & -0.20 & Truncated normal \\
$r$       & 0.50 & 40 & 0.20  & 0.80  & Truncated normal \\
$\mu_u$   & 6.0  & 40 & 2.40  & 9.60  & Truncated normal \\
\hline
\end{tabular}
\end{table}

\subsubsection{Structural Response}

\noindent
For the IDA calculations, a suite of 30 ordinary ground-motion records was
used (Table~\ref{tab:ground_motions}). The records represent a relatively
narrow earthquake scenario, with moment magnitudes between $6.5$ and $6.9$
and moderate source-to-site distances. The motions were recorded primarily
on firm soil and were selected without strong directivity effects. For each
record, IDA consists of repeated nonlinear response-history analyses in which
the ground motion is progressively scaled over increasing intensity levels.
\begin{table}[!htb]
\centering
\caption{Suite of 30 ground-motion records used in the incremental dynamic analyses.}
\label{tab:ground_motions}
\resizebox{\textwidth}{!}{%
\begin{tabular}{cllccccc}
\hline
No. & Event & Station & Component ($^\circ$) & Soil & $M$ & $R$ (km) & PGA (g) \\
\hline
1  & Loma Prieta, 1989       & Agnews State Hospital          & 090 & C,D & 6.9 & 28.2 & 0.159 \\
2  & Northridge, 1994        & LA, Baldwin Hills              & 090 & B,B & 6.7 & 31.3 & 0.239 \\
3  & Imperial Valley, 1979   & Compuertas                     & 285 & C,D & 6.5 & 32.6 & 0.147 \\
4  & Imperial Valley, 1979   & Plaster City                   & 135 & C,D & 6.5 & 31.7 & 0.057 \\
5  & Loma Prieta, 1989       & Hollister Diff. Array          & 255 & --,D & 6.9 & 25.8 & 0.279 \\
6  & San Fernando, 1971      & LA, Hollywood Stor. Lot        & 180 & C,D & 6.6 & 21.2 & 0.174 \\
7  & Loma Prieta, 1989       & Anderson Dam Downstrm          & 270 & B,D & 6.9 & 21.4 & 0.244 \\
8  & Loma Prieta, 1989       & Coyote Lake Dam Downstrm       & 285 & B,D & 6.9 & 22.3 & 0.179 \\
9  & Imperial Valley, 1979   & El Centro Array \#12           & 140 & C,D & 6.5 & 18.2 & 0.143 \\
10 & Imperial Valley, 1979   & Cucapah                        & 085 & C,D & 6.5 & 23.6 & 0.309 \\
11 & Northridge, 1994        & LA, Hollywood Storage FF       & 360 & C,D & 6.7 & 25.5 & 0.358 \\
12 & Loma Prieta, 1989       & Sunnyvale Colton Ave           & 270 & C,D & 6.9 & 28.8 & 0.207 \\
13 & Loma Prieta, 1989       & Anderson Dam Downstrm          & 360 & B,D & 6.9 & 21.4 & 0.240 \\
14 & Imperial Valley, 1979   & Chihuahua                      & 012 & C,D & 6.5 & 28.7 & 0.270 \\
15 & Imperial Valley, 1979   & El Centro Array \#13           & 140 & C,D & 6.5 & 21.9 & 0.117 \\
16 & Imperial Valley, 1979   & Westmoreland Fire Station      & 090 & C,D & 6.5 & 15.1 & 0.074 \\
17 & Loma Prieta, 1989       & Hollister South \& Pine        & 000 & --,D & 6.9 & 28.8 & 0.371 \\
18 & Loma Prieta, 1989       & Sunnyvale Colton Ave           & 360 & C,D & 6.9 & 28.8 & 0.209 \\
19 & Superstition Hills, 1987& Wildlife Liquefaction Array    & 090 & C,D & 6.7 & 24.4 & 0.180 \\
20 & Imperial Valley, 1979   & Chihuahua                      & 282 & C,D & 6.5 & 28.7 & 0.254 \\
21 & Imperial Valley, 1979   & El Centro Array \#13           & 230 & C,D & 6.5 & 21.9 & 0.139 \\
22 & Imperial Valley, 1979   & Westmoreland Fire Station      & 180 & C,D & 6.5 & 15.1 & 0.110 \\
23 & Loma Prieta, 1989       & Halls Valley                    & 090 & C,C & 6.9 & 31.6 & 0.103 \\
24 & Loma Prieta, 1989       & WAHO                            & 000 & --,D & 6.9 & 16.9 & 0.370 \\
25 & Superstition Hills, 1987& Wildlife Liquefaction Array    & 360 & C,D & 6.7 & 24.4 & 0.200 \\
26 & Imperial Valley, 1979   & Compuertas                     & 015 & C,D & 6.5 & 32.6 & 0.186 \\
27 & Imperial Valley, 1979   & Plaster City                   & 045 & C,D & 6.5 & 31.7 & 0.042 \\
28 & Loma Prieta, 1989       & Hollister Diff. Array          & 165 & --,D & 6.9 & 25.8 & 0.269 \\
29 & San Fernando, 1971      & LA, Hollywood Stor. Lot        & 090 & C,D & 6.6 & 21.2 & 0.210 \\
30 & Loma Prieta, 1989       & WAHO                            & 090 & --,D & 6.9 & 16.9 & 0.638 \\
\hline
\end{tabular}%
}

\vspace{2mm}
\begin{flushleft}
\footnotesize
Component denotes the horizontal ground-motion component;
Soil denotes the USGS/Geomatrix soil classification;
$M$ is moment magnitude;
$R$ is the closest distance to fault rupture;
PGA is the unscaled peak ground acceleration. Adopted from \cite{vamvatsikos2010incremental}.
\end{flushleft}
\end{table}
The intensity measure (IM) is the 5\%-damped first-mode spectral acceleration,
\begin{equation}
    IM = S_a(T_1,5\%),
\end{equation}
where $T_1=2.35$~s is the fundamental period of the reference LA9 structure.
The corresponding engineering demand parameter (EDP) is the maximum
interstory drift, $EDP = D_{\max},$
consistent with the emphasis of the present study on deformation-sensitive
structural and nonstructural response. Each record therefore defines an IDA
trajectory relating $S_a(T_1,5\%)$ to $D_{\max}$. 

The same suite of 30 ground-motion records was used for every structural
realization, with $S_a(T_1,5\%)$ as the intensity measure and maximum
interstory drift $D_{\max}$ as the engineering demand parameter. For each
realization of $\mathbf{x}_s^i$, $i=1,\ldots,160$, the resulting 30
single-record IDAs were available from previous nonlinear analyses performed
using the hunt-and-fill algorithm \cite{vamvatsikos2004applied}. 
Each single-record IDA required approximately 12--14 nonlinear analyses, corresponding to approximately 360--420 nonlinear dynamic analyses per structural realization and approximately 57,600--67,200 analyses over the full 160-realization dataset. On a conventional laptop equipped with an 8-core Apple M1 Pro processor and 16~GB of unified memory,  one complete 30-record adaptive
IDA required on average approximately 42.3~min per structural realization,
with individual runtimes ranging from 40.5 to 43.1~min. The
160 realization dataset corresponds to approximately 113~h of aggregate
computation, or approximately 38~h of wall-clock time when three realizations
are processed concurrently.

\begin{figure}[!htb]
\centering
\begin{subfigure}[t]{0.32\textwidth}
    \centering
    \includegraphics[width=\textwidth]{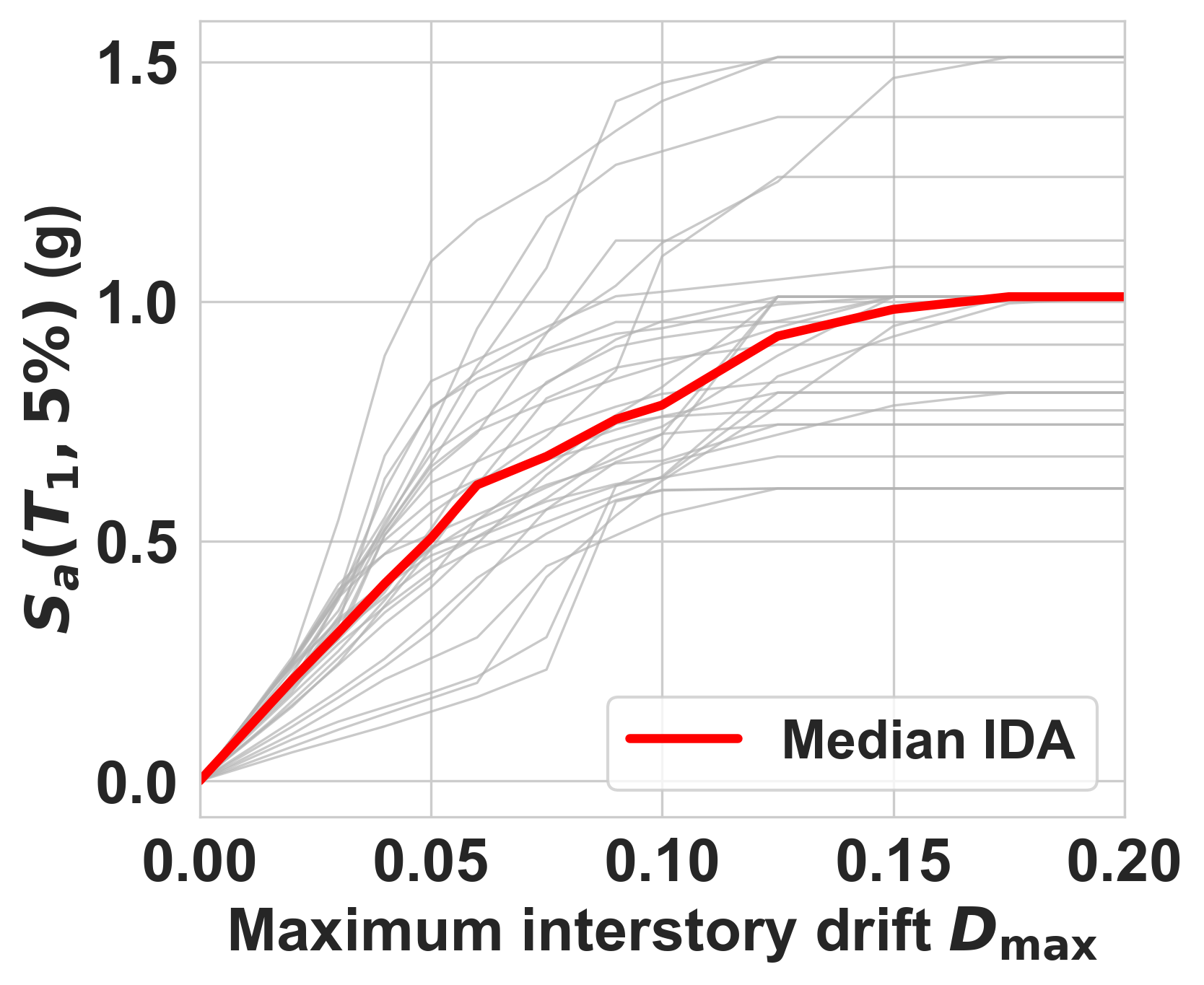}
\caption{Thirty ground-motion IDA curves for one structural realization, illustrating record-to-record variability; the red curve denotes their pointwise median.}
    \label{fig:ida_single_structure}
\end{subfigure}
\hfill
\begin{subfigure}[t]{0.32\textwidth}
    \centering
    \includegraphics[width=\textwidth]{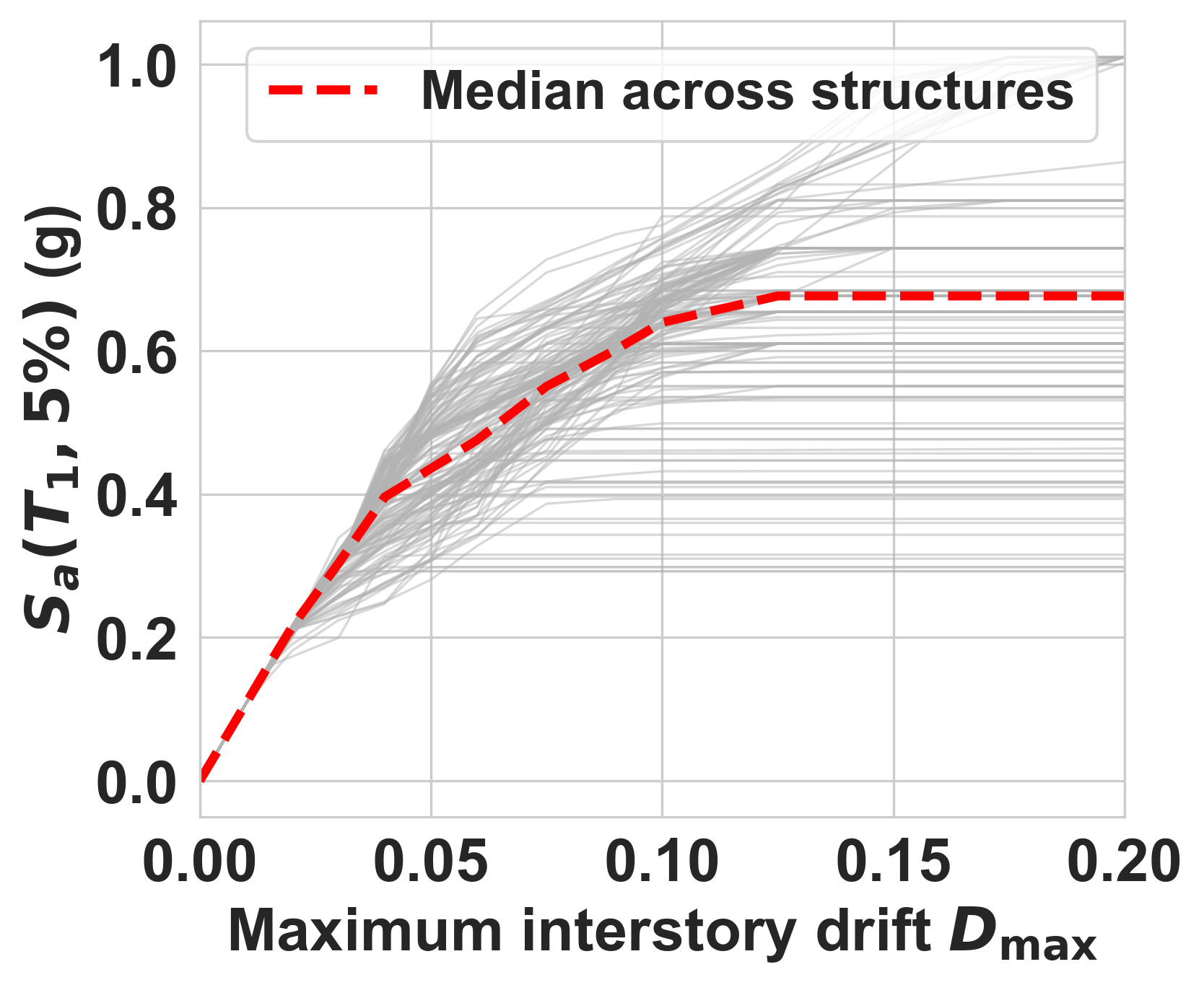}
\caption{IDA responses of all 150 structural realizations for one fixed ground-motion record, illustrating structural variability; the dashed red curve denotes their pointwise median.}
    \label{fig:ida_single_record}
\end{subfigure}
\hfill
\begin{subfigure}[t]{0.32\textwidth}
    \centering
    \includegraphics[width=\textwidth]{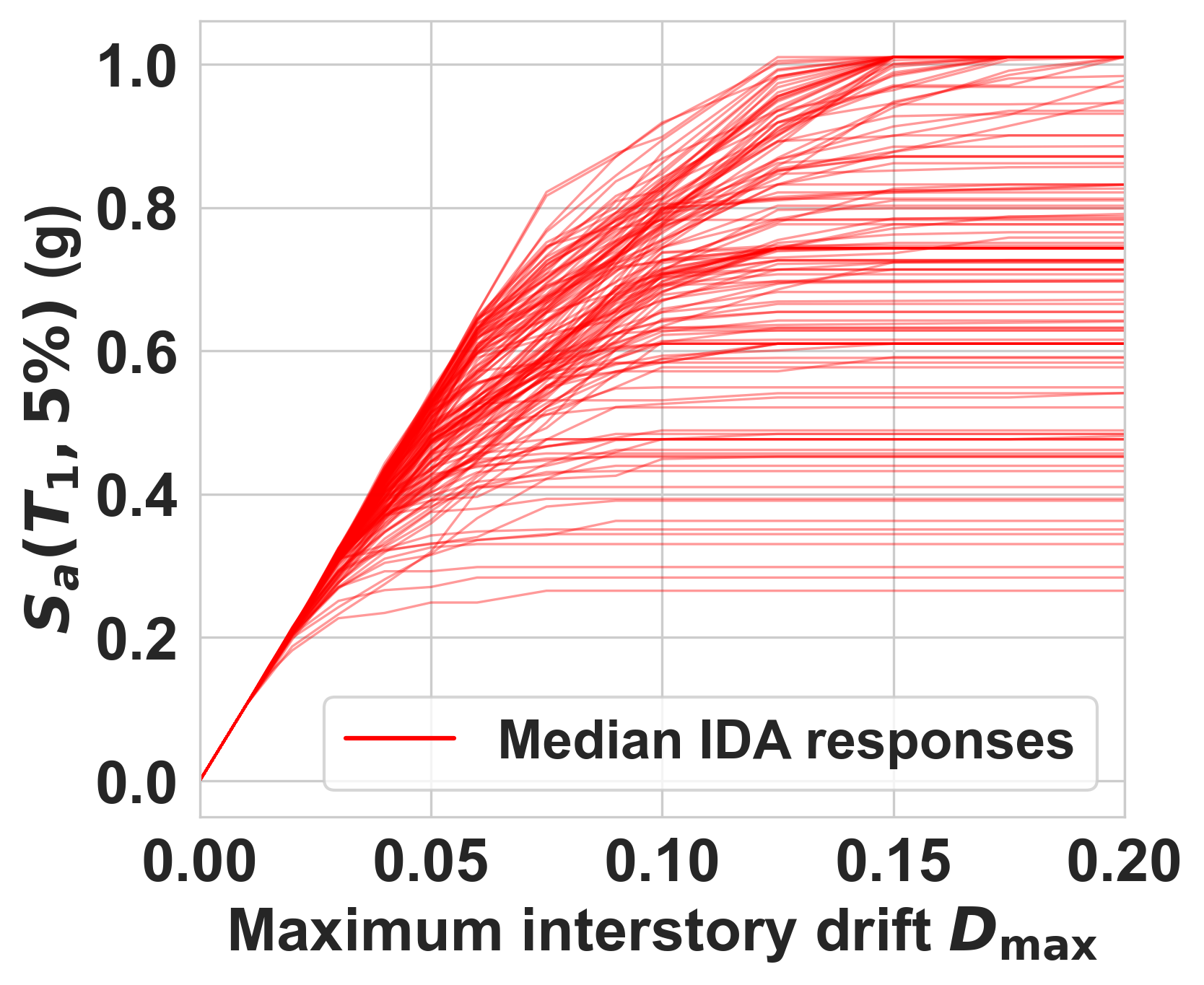}
\caption{Median IDA curves for all 150 structural realizations, obtained by taking the pointwise median across the 30 ground-motion records.}
    \label{fig:ida_all_medians}
\end{subfigure}
\caption{Illustration of the variability represented in the IDA dataset. 
(a) Record-to-record variability for a fixed structural realization; 
(b) structural-parameter variability for a fixed ground-motion record; and 
(c) representative median IDA responses across the full set of structural realizations.}
\label{fig:ida_variability}
\end{figure}

The present study constructs a fixed-dimensional
response representation for each structural realization. Each record-level IDA
curve is first converted to a first-passage monotone relation $S_a(D)$ and
interpolated onto the common 18-point drift grid. The 30 processed responses
associated with a given structural realization are then aggregated pointwise by
taking the median spectral acceleration at each drift coordinate. The resulting
representative response is $\mathbf{x}_r
=
\widetilde{\mathbf{S}}_a$. 
Figure~\ref{fig:ida_variability} illustrates the two sources of variability represented in the IDA dataset and the resulting median structural responses. Figure~\ref{fig:ida_single_structure} fixes one structural realization and shows its responses to the 30 ground-motion records, thereby illustrating the record-to-record variability in the IDA response. The red curve denotes the pointwise median across the 30 records. Figure~\ref{fig:ida_single_record} instead fixes one ground-motion record and shows the responses of all 160 structural realizations, highlighting the variability induced by changes in the structural parameters; the dashed red curve denotes the pointwise median across these realizations. Finally, Fig.~\ref{fig:ida_all_medians} shows the representative median IDA curve associated with each of the 160 structural realizations, where each curve is obtained by taking the pointwise median across the 30 ground motion records. 

\section{Results}
\label{sec:results}

\noindent
We evaluate the proposed framework using a nine-story steel moment-resisting frame subjected to earthquake ground motions (see Section \ref{sec:LA9}). Structural uncertainty is represented through six parameters describing the nonlinear strength and deformation characteristics of the structural components: the normalized yield moment ($aM_y$), strain-hardening ratio ($a_h$), capping ductility ($\mu_c$), post-capping slope ($a_c$), residual-strength ratio ($r$), and ultimate ductility ($\mu_u$). For each structural realization, Incremental Dynamic Analysis (IDA) is performed using 30 earthquake ground-motion records (see Table \ref{tab:ground_motions}), progressively scaled to characterize structural performance from the elastic to the highly nonlinear regime. 
The resulting Incremental Dynamic Analysis (IDA) curves describe the relationship between ground-motion intensity and structural deformation. Ground-motion intensity is quantified by the 5\%-damped spectral acceleration $S_a(T_1,5\%)$, defined as the peak acceleration response of an equivalent linear single-degree-of-freedom oscillator with a natural period equal to the fundamental period $T_1$ of the structure and a damping ratio of 5\%. Structural deformation is measured by the maximum interstory drift ratio $D_{\max}$, defined as the maximum relative horizontal displacement between two consecutive floors, normalized by the corresponding story height, over the duration of the earthquake excitation. Each representative IDA curve is evaluated at 18 prescribed drift levels, providing a fixed-dimensional description of how the ground-motion intensity required to reach a given deformation level evolves from the elastic to the highly nonlinear structural response regime.

The dataset comprises 160 structural realizations, of which 151 are used for training and 9 are reserved for validation. The validation cases are selected using a maximin distance criterion in the standardized six-dimensional structural-parameter space to obtain a diverse set of parameter configurations. The Diffusion Maps algorithm (see Appendix \ref{sec:DMaps}) is applied to the standardized response vectors to identify a compact nonlinear representation of the seismic-response manifold. This reduction exploits the intrinsic geometric structure of the IDA trajectories, allowing their dominant variability to be represented by a small number of non-redundant coordinates. In contrast, the six structural parameters are retained explicitly in the joint generative representation to preserve their physical meaning and enable direct conditioning on individual parameters. This hybrid representation reduces the dimensionality of the generative learning problem while supporting both forward probabilistic prediction and inverse inference of structural properties from observed seismic responses.

\subsection{Identification and Validation of the Response Manifold}
To identify a compact representation of the seismic-response manifold, Diffusion Maps are applied to the standardized 18-dimensional representative IDA responses. The method constructs a similarity kernel that measures the proximity of response trajectories, with a bandwidth parameter $\epsilon_r$ controlling the scale over which neighboring responses are considered similar. Here, the bandwidth is selected as $\epsilon_r
=
10\times
\operatorname{median}_{i<j}
\left\|
\mathbf{x}_{r,i}^{(\mathrm{std})}
-
\mathbf{x}_{r,j}^{(\mathrm{std})}
\right\|^2,$
where $\mathbf{x}_{r,i}^{(\mathrm{std})}$ denotes the standardized response vector for the $i$th training realization. This gives $\epsilon_r=172.72$ for the present training dataset. The first 15 nontrivial diffusion eigenpairs are computed to examine the intrinsic structure of the response manifold. Because some diffusion eigenvectors may describe higher-order harmonics of coordinates already identified, retaining all of them would introduce redundant information. We therefore apply the local-regression criterion of \cite{dsilva2018parsimonious}, which evaluates whether each candidate diffusion coordinate can be locally approximated using the preceding coordinates. The resulting normalized regression residual $r_k$ measures the extent to which the $k$th coordinate contains information not explained by the previously considered coordinates. Larger residuals indicate potentially independent geometric directions, whereas smaller residuals suggest harmonic redundancy.

\begin{figure}[!htb]
    \centering

    \begin{subfigure}[t]{0.38\textwidth}
        \centering
        \includegraphics[width=\textwidth]{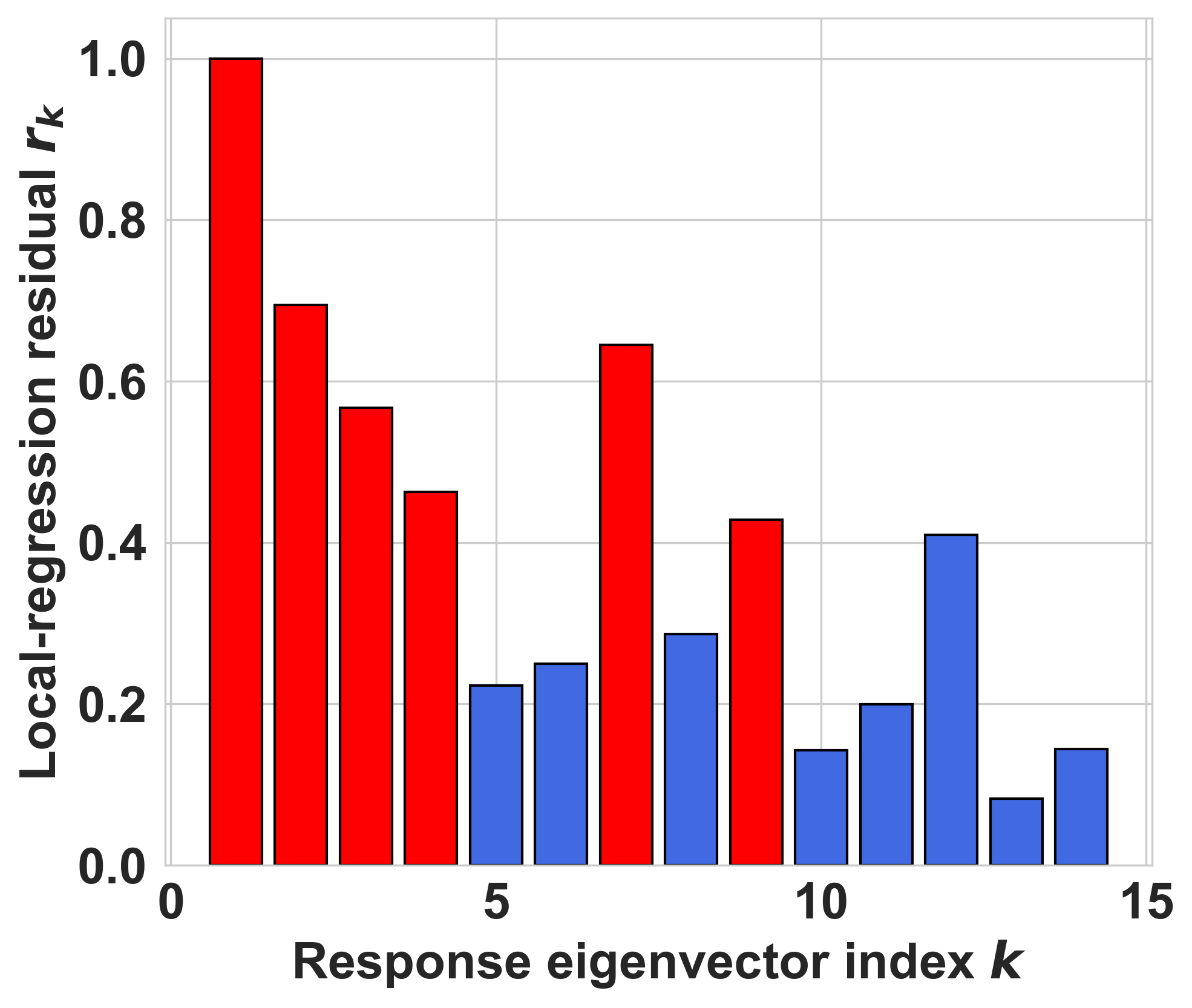}
        \caption{
        Normalized local-regression residuals $r_k$ from the response-only
        Diffusion Maps analysis. Red bars indicate the six non-harmonic
        eigenvectors retained in the response representation.
        }
        \label{fig:ida_DMAPS_res}
    \end{subfigure}
    \hfill
    \begin{subfigure}[t]{0.58\textwidth}
        \centering
        \includegraphics[width=\textwidth]{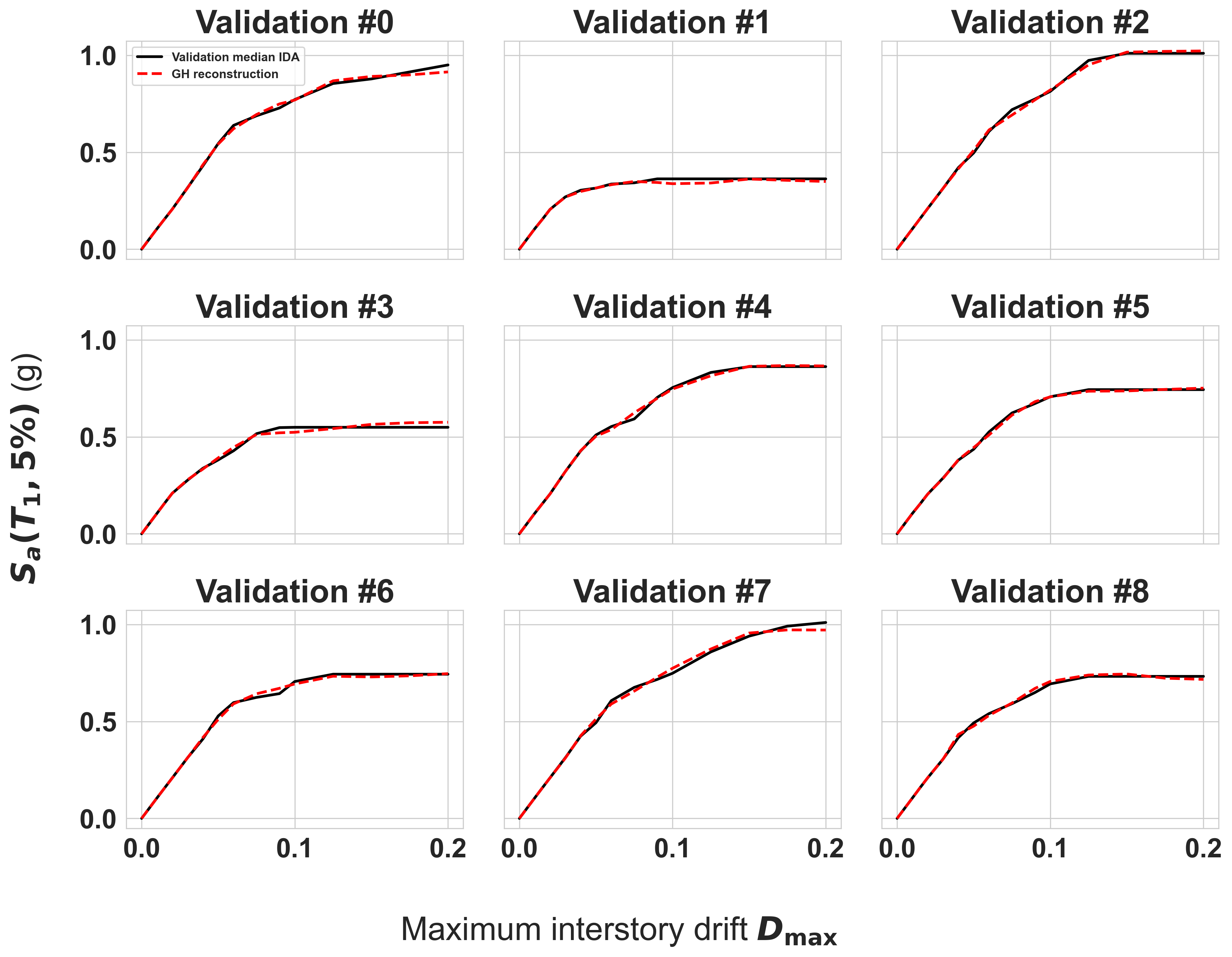}
        \caption{
        Geometric-Harmonics reconstruction of the nine held-out representative
        IDA responses. Solid black curves denote the held-out median IDA
        responses and dashed red curves denote their reconstructions.
        }
        \label{fig:GH_validation}
    \end{subfigure}

    \caption{
    Selection and validation of the response-manifold representation.
    (a) Non-harmonic Diffusion Maps coordinate selection using local-regression
    residuals. (b) Held-out reconstruction of the 18-dimensional representative
    IDA responses using the 6 retained coordinates and Geometric Harmonics.
    }
    \label{fig:response_manifold_validation}
\end{figure}

Figure~\ref{fig:ida_DMAPS_res} shows the normalized residuals for the evaluated diffusion coordinates. Six non-harmonic coordinates are retained (red bars), providing a compact representation of the dominant geometric variability in the IDA responses. The remaining coordinates are treated as predominantly redundant and are excluded from the subsequent generative learning procedure. The original parameter--response vector is $\mathbf{x}
=
\left(
\mathbf{x}_s,
\mathbf{x}_r
\right)
=
\left(
\mathbf{x}_s,
\widetilde{\mathbf{S}}_a
\right)
\in
\mathbb{R}^{6+18}
=
\mathbb{R}^{24},$
where $\mathbf{x}_s$ contains the six uncertain structural parameters and $\mathbf{x}_r=\widetilde{\mathbf{S}}_a$ denotes the representative IDA response. Following dimensionality reduction, the joint generative state is defined as
\[
\mathbf{z}
=
\left(
\mathbf{x}_s^{(\mathrm{std})},
\boldsymbol{\phi}_r
\right)
\in
\mathbb{R}^{6+6}
=
\mathbb{R}^{12},
\]
where $\boldsymbol{\phi}_r$ contains the six retained Diffusion Maps coordinates. This representation reduces the dimension of the generative learning problem while retaining the structural parameters explicitly for physical interpretation and conditional inference. To reconstruct the original seismic responses from the reduced coordinates, we employ Geometric Harmonics (GH) (see Section \ref{sec:GH}), a kernel-based method that approximates the inverse mapping from the learned manifold coordinates to the original response space. GH is applied only to the response variables, lifting the six retained diffusion coordinates back to the 18-dimensional drift representation. The GH kernel bandwidth is set to 2.5 times the median squared pairwise distance in the response latent space.

The accuracy of this reduced representation is evaluated using the nine held-out structural realizations, which were not used to construct the response manifold or its reconstruction map. Figure~\ref{fig:GH_validation} compares their original representative IDA curves with those reconstructed from the six retained coordinates. The overall root-mean-square error (RMSE) is $0.0106$~g, with individual errors ranging from $0.0046$ to $0.0147$~g. Note that the RMSE is expressed in units of gravitational acceleration ($g$), since the reconstruction error is evaluated in terms of spectral acceleration at the 18 prescribed interstory drift levels. These results indicate that the six-dimensional manifold representation preserves the principal characteristics of the seismic-response trajectories with limited reconstruction error, supporting its use in the subsequent joint generative learning procedure.

\subsection{Physical Interpretation of the Manifold and Structural Parameter Recoverability}

\begin{figure}[!htb]
\centering
\includegraphics[width=1.0\textwidth]{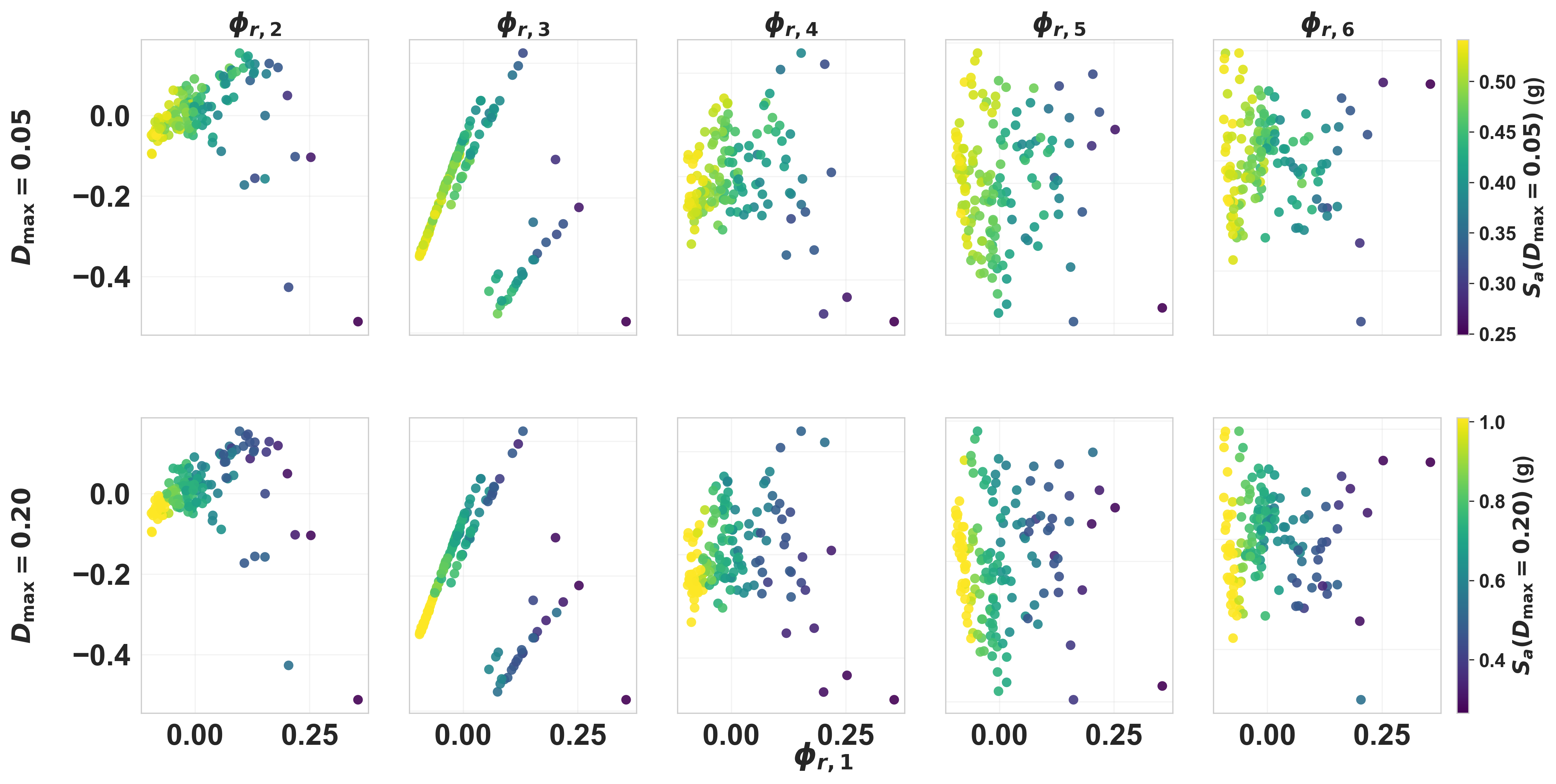}
\caption{Pairwise projections of the retained response
Diffusion Maps coordinates, colored by the representative
spectral acceleration at selected maximum interstory
drift levels.}
\label{fig:ida_DMAPS}
\end{figure}

Having established that the six-dimensional Diffusion Maps
representation accurately reconstructs the representative
IDA responses, we next investigate the physical information
encoded in the learned coordinates and their relationship
to the uncertain structural parameters. Figure~\ref{fig:ida_DMAPS} shows pairwise projections
of the retained Diffusion Maps coordinates, colored
by the spectral acceleration $S_a$ at selected maximum
interstory drift levels. The response values vary
smoothly across the embedding, indicating that nearby
points tend to correspond to IDA trajectories with
similar response characteristics. Moreover, the
organization of spectral acceleration changes across
drift levels, suggesting that the retained coordinates
capture variations in the shape and evolution of the
nonlinear IDA response rather than merely representing
a single response amplitude. These observations
provide qualitative evidence that the learned manifold
preserves physically meaningful variations in seismic
performance.

An important question is whether the learned response
manifold contains sufficient information to infer the
underlying structural properties from observed seismic
performance. Parameter identifiability refers to the
ability to uniquely infer the parameters of a model
from observations of its behavior. More formally,
identifiability concerns whether the mapping from
model parameters to observable responses is injective,
such that distinct parameter configurations produce
distinct responses and an inverse mapping can be defined.
When this condition is not satisfied, different parameter
combinations may produce the same observable behavior,
leading to structural non-identifiability. In the present setting, we consider the mapping from
the six uncertain structural parameters,
$\boldsymbol{\theta}
=(aM_y,a_h,\mu_c,a_c,r,\mu_u)$,
to the six retained Diffusion Maps coordinates,
$\boldsymbol{\phi}_r
=(\phi_{r,1},\ldots,\phi_{r,6})$,
which provide a compact representation of the
seismic-response trajectories $\boldsymbol{\phi}_r
=
\mathcal{F}(\boldsymbol{\theta}).$
From an analytical perspective, local identifiability
can be investigated through the linearization of
this mapping, characterized by its Jacobian,
\[
\mathbf{J}(\boldsymbol{\theta})
=
\frac{\partial\mathcal{F}(\boldsymbol{\theta})}
{\partial\boldsymbol{\theta}}.
\]
For a continuously differentiable mapping, a
full-rank Jacobian provides a sufficient condition
for local invertibility. Conversely, small singular
values indicate parameter directions along which
the observable response changes only weakly.
Such directions may result in large parametric
uncertainties and practical difficulties in
distinguishing different structural configurations.
Importantly, even when a mapping is locally
invertible, this does not guarantee global
uniqueness of the inverse solution. Beyond this local analytical perspective,
manifold learning methods offer a data-driven
approach to examining the relationship between
model parameters and observable behavior.
In particular, the intrinsic geometry of the
response manifold may reveal that certain
parameters, or combinations of parameters,
are more strongly expressed in the observable
responses than others. These identifiable
combinations need not coincide with the
individual physical parameters used to
formulate the structural model.

Rather than explicitly computing the Jacobian
or the Fisher Information Matrix \cite{rothenberg1971identification, brouwer2018underlying}, we investigate
parameter recoverability directly from the
learned response manifold. We first examine
how individual structural parameters vary across
the Diffusion Maps embedding. We then
attempt to reconstruct each parameter from the
six retained diffusion coordinates using several
nonlinear regression methods, including kernel
ridge regression, support vector regression,
random forests, extremely randomized trees,
and gradient boosting. Predictive performance is evaluated using
nested five-fold cross-validation, with
hyperparameters selected exclusively within
the inner training folds. The resulting
out-of-sample coefficient of determination,
$R^2$, quantifies how effectively each
structural parameter can be recovered from
the learned response representation.
Higher $R^2$ values indicate stronger empirical
recoverability, whereas lower values may
reflect weak observable dependence, confounding
between parameters, or limitations associated
with finite data and regression accuracy. This analysis provides a data-driven assessment
of parameter recoverability rather than a
formal proof of structural identifiability.
Importantly, it also motivates the investigation
of whether physically meaningful combinations
of backbone parameters may be more
recoverable than the individual parameters
themselves, as examined in the subsequent
analysis of post-capping degradation.

\begin{figure}[!htb]
\centering

\begin{subfigure}[t]{0.48\textwidth}
    \centering
    \includegraphics[width=\textwidth]{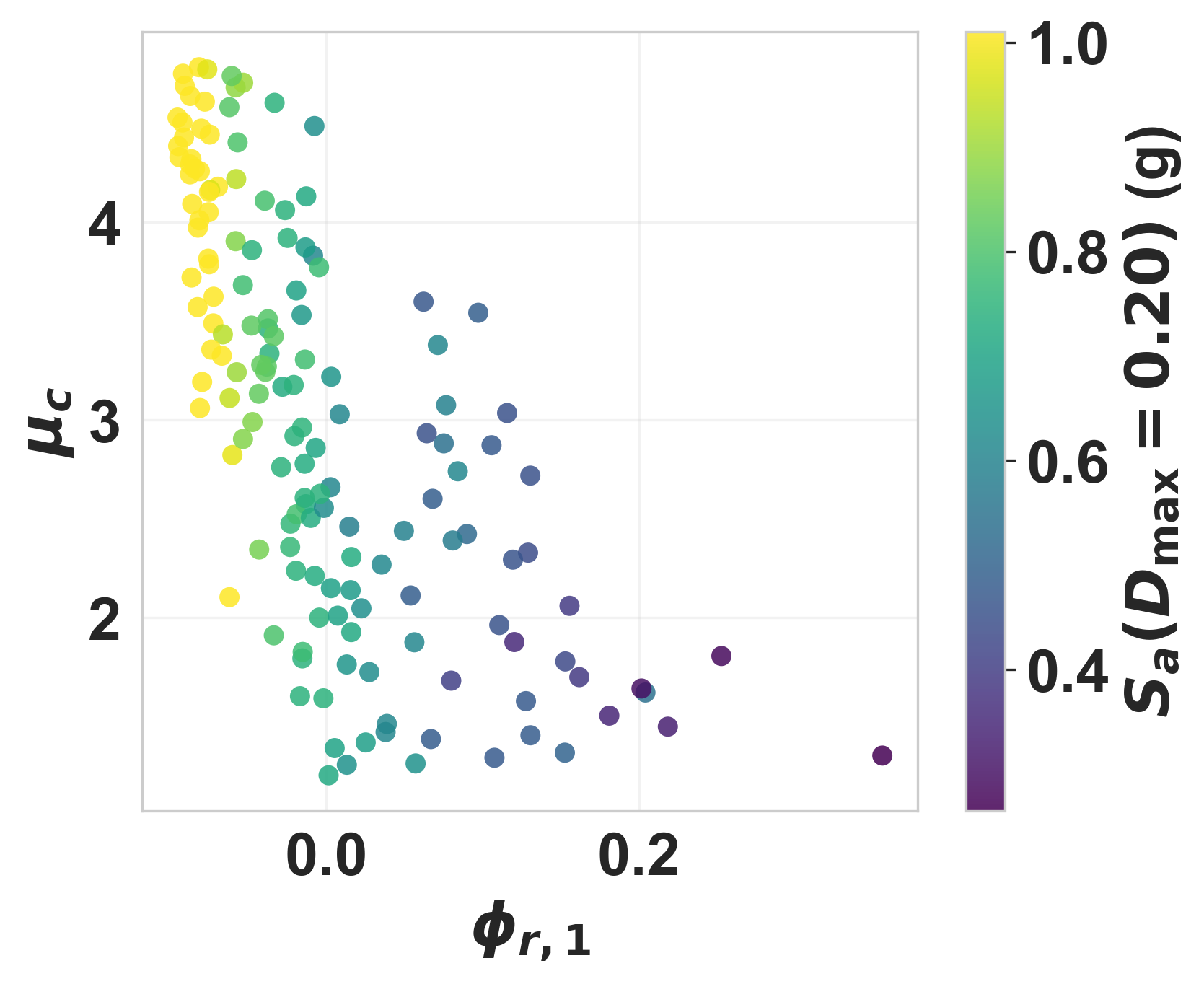}
    \caption{}
    \label{fig:muc_phi1}
\end{subfigure}
\hfill
\begin{subfigure}[t]{0.48\textwidth}
    \centering
    \includegraphics[width=\textwidth]{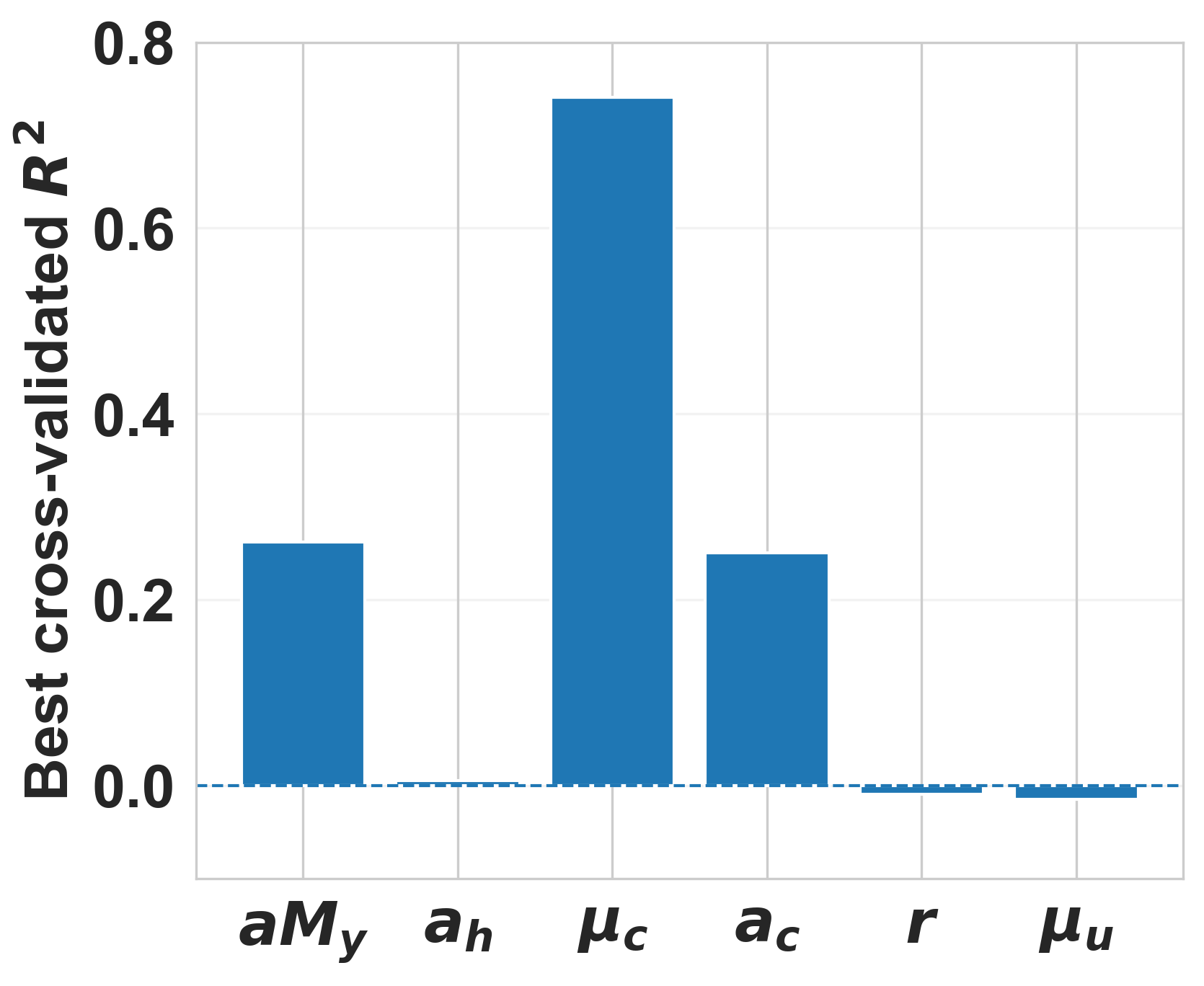}
    \caption{}
    \label{fig:param_recovery}
\end{subfigure}

\caption{Physical interpretation of the learned IDA-response manifold. 
Panel (a) shows that the leading diffusion coordinate is strongly associated with the capping rotation $\mu_c$ and with the spectral intensity required to reach large drift. 
Panel (b) quantifies the recoverability of the six backbone parameters and shows that $\mu_c$ is the parameter most strongly encoded in the learned manifold.}
\label{fig:physical_manifold}

\end{figure}

Figure~\ref{fig:physical_manifold}(a) illustrates the clearest relationship identified in this analysis. The capping rotation $\mu_c$ varies systematically with the leading diffusion coordinate $\phi_{r,1}$: larger values of $\mu_c$ are predominantly located at smaller values of $\phi_{r,1}$, whereas smaller values of $\mu_c$ occur toward larger $\phi_{r,1}$. The points are additionally colored by $S_a$ evaluated at $D_{\max}=0.20$, revealing that the same region associated with larger $\mu_c$ also corresponds to larger spectral acceleration. Thus, $\phi_{r,1}$ is not simply an abstract latent coordinate, but organizes the IDA responses along a direction that is simultaneously related to a physically meaningful backbone parameter and to the intensity required to attain large drift. This interpretation is consistent with the role of $\mu_c$ in controlling the transition from the hardening branch to post-capping degradation, a feature that can have a pronounced influence on the nonlinear IDA response.

The quantitative reconstruction results in Fig.~\ref{fig:physical_manifold}(b) support this interpretation. Across the different nonlinear regressors, $\mu_c$ was consistently the most recoverable structural parameter, with cross-validated $R^2$ values of approximately $0.72$--$0.75$. In contrast, $aM_y$ and $a_c$ showed only modest recoverability, with best $R^2$ values of approximately $0.26$ and $0.25$, respectively, while $a_h$, $r$, and $\mu_u$ produced $R^2$ values close to zero. The consistency of this ranking across  different regression models indicates that  their effects appear to be weakly expressed or non-unique within the median IDA responses used to construct the manifold.

An interesting result is that the post-capping
slope parameter $a_c$ is more strongly recoverable from the learned response
coordinates than either the hardening parameter $a_h$ or the residual-strength
parameter $r$. This ranking should not be interpreted as a conventional
forward-sensitivity measure of the structural model. Rather, it quantifies
how distinctly variations in each parameter are expressed in the median IDA
responses used to construct the manifold. The stronger recoverability of
$a_c$ suggests that post-capping degradation leaves a more distinguishable
signature over the response range sampled by the IDA curves, whereas the
effects of $a_h$ and $r$ may be more localized or partially confounded with
variations in other parameters. The result therefore highlights the
distinction between physical influence on the nonlinear system and
identifiability from the selected response observable.

\subsection{Generative learning of the compact state space}
\noindent
The score-based diffusion modeling framework (see Section \ref{sec:SGM}) was applied directly to the 12-dimensional joint generative state formed by the six standardized structural parameters and the six retained Diffusion Maps response coordinates. The probability-flow Ordinary Differential Equation (ODE) associated with the forward diffusion process was
integrated backward from $t=1$ to $t=0$ using 1000 uniform explicit-Euler
steps. At each step, the empirical score was evaluated using the fixed
training reference ensemble.  A total of 10,000 standard normal samples were transported through this probability-flow ODE. Generated realizations were retained when all six structural parameters remained within their prescribed bounds and all response variables were finite and non-negative. This produced 9,850 admissible realizations.

\begin{figure}[!htb]
\centering
\includegraphics[width=0.92\textwidth]{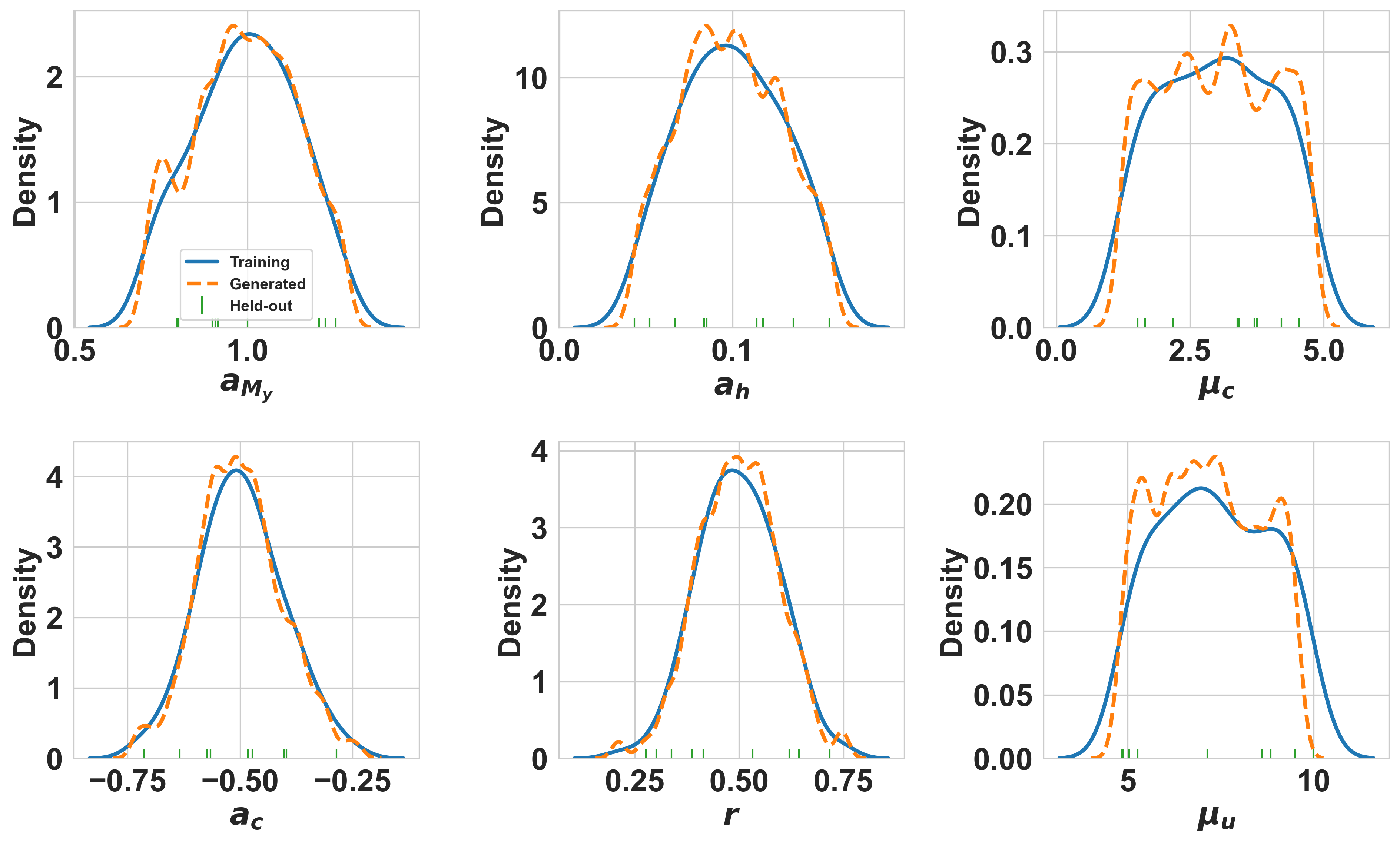}
\caption{Marginal distributions of the six structural parameters. Solid curves correspond to the training samples, dashed curves correspond to the admissible generated samples, and vertical marks denote the nine held-out validation cases.}
\label{fig:param_marginals}
\end{figure}

To assess the accuracy of the learned joint distribution, we compare the generated structural parameters and seismic
responses with those in the original simulation dataset. In addition to examining their marginal distributions, we investigate whether the generated samples reproduce the dependence between structural properties and seismic
performance. We then evaluate the model's ability to generate representative IDA responses consistent with the held-out validation cases and to capture the changes
in response distributions induced by conditioning on
specific structural parameters.
Figure~\ref{fig:param_marginals} shows that the generated samples reproduce the training marginals closely for all six structural parameters. As an additional diagnostic, particular attention was given to the lower tail of the capping ductility parameter \(\mu_c\), which was found during model development to be especially sensitive to generative smoothing. The 5th percentile is 1.378 in the training ensemble and 1.364 in the admissible generated ensemble. The corresponding lower-tail probabilities at \(\mu_c=1.50\), 1.60, 1.80, and 2.00 are, respectively, 0.086, 0.106, 0.166, and 0.212 for the training data and 0.090, 0.111, 0.168, and 0.216 for the generated data. These results indicate that the generative model preserves this sensitive lower-tail behavior closely.

\begin{figure}[!htb]
    \centering

    \begin{subfigure}[t]{0.48\textwidth}
        \centering
        \includegraphics[width=\textwidth]{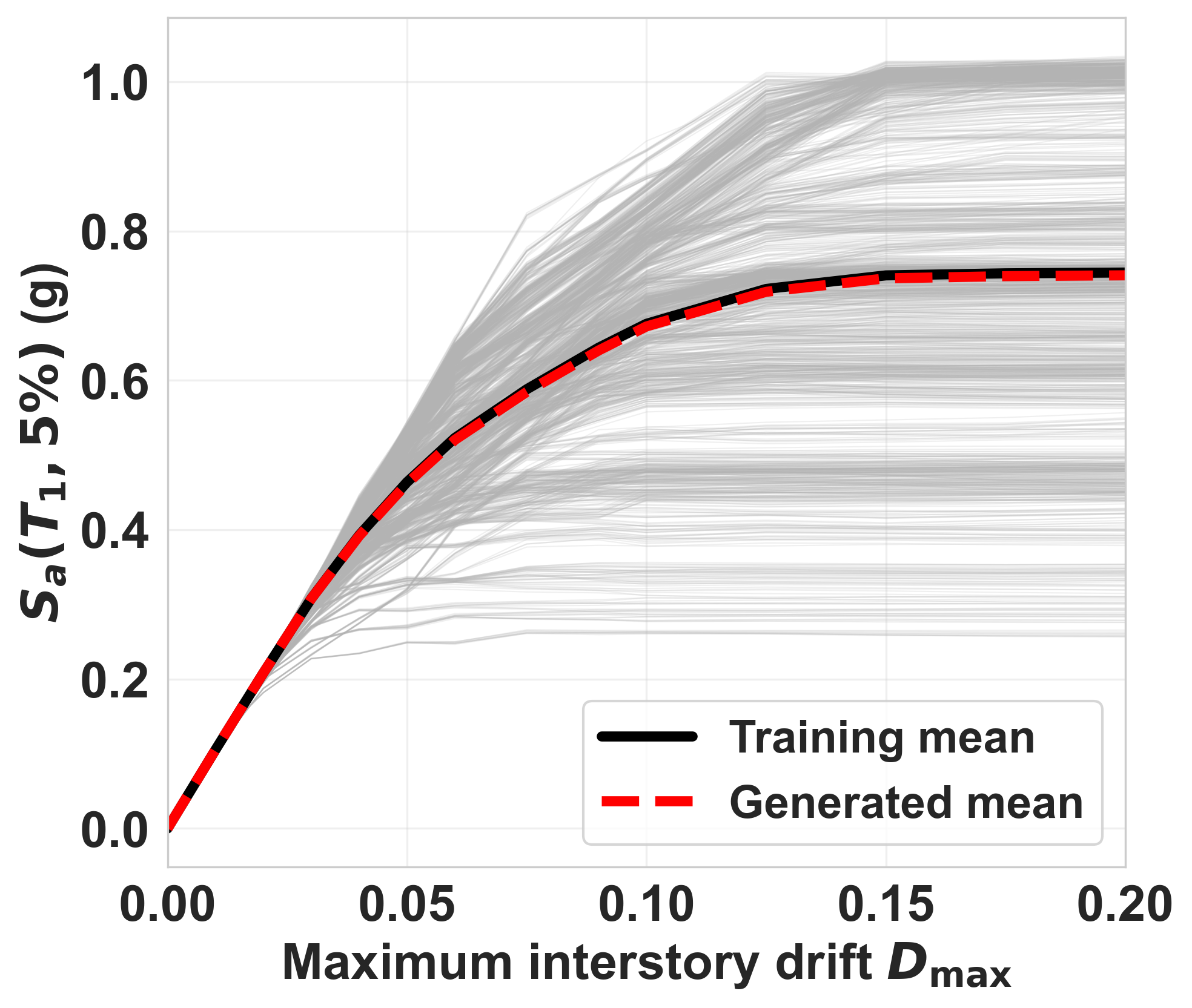}
        \caption{
        Generated representative IDA responses. Gray lines denote 1,600
        admissible generated responses, the red dashed line denotes their mean,
        and the black solid line denotes the mean response of the training
        structural realizations.
        }
        \label{fig:ida_DMAPS_curves}
    \end{subfigure}
    \hfill
    \begin{subfigure}[t]{0.48\textwidth}
        \centering
        \includegraphics[width=\textwidth]{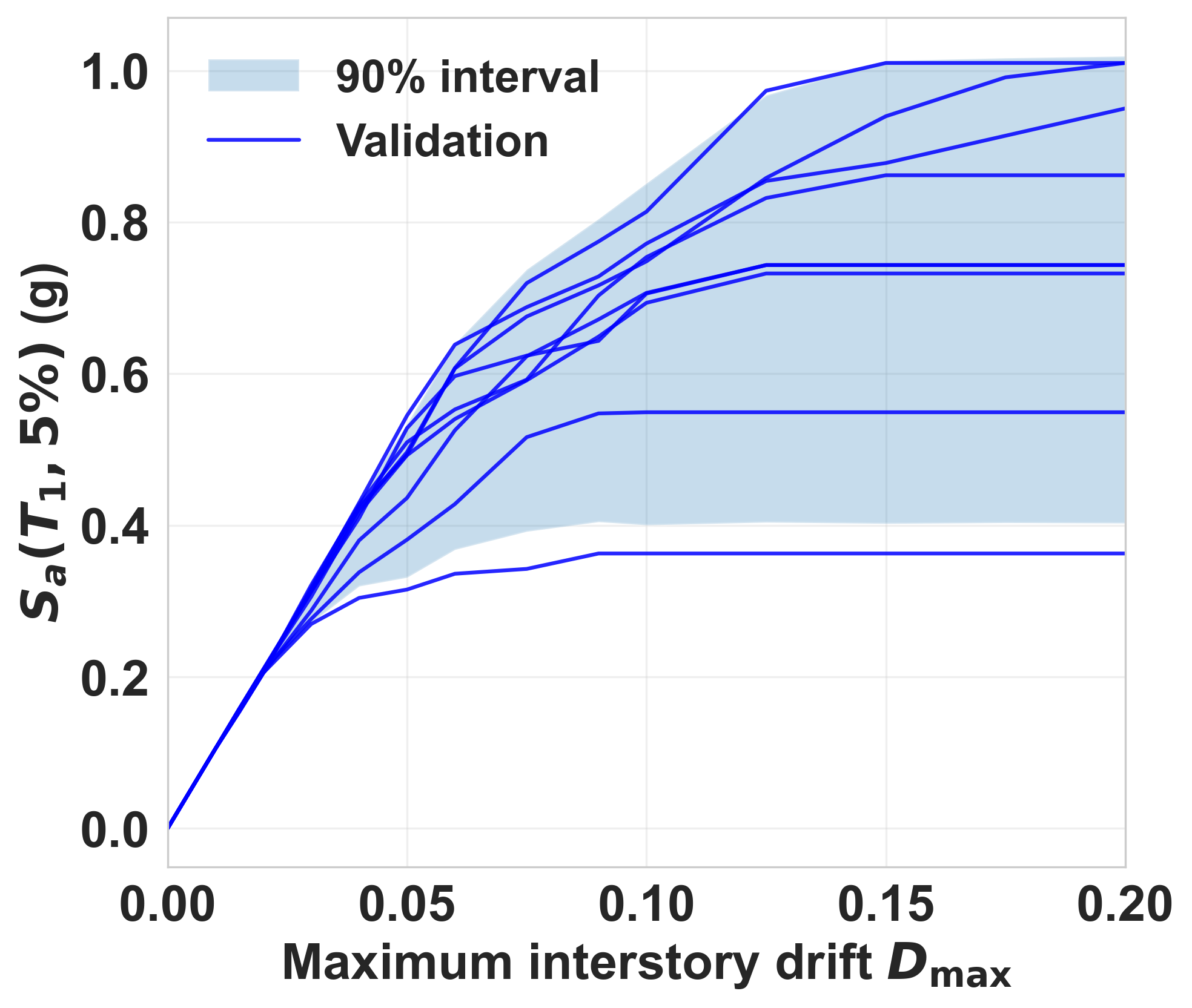}
        \caption{
        Nine held-out representative IDA responses together with the
        5--95\% interval of the admissible generated response distribution.
        }
        \label{fig:validation_envelope}
    \end{subfigure}

    \caption{
    Unconditional generative performance and held-out validation.
    (a) Representative IDA responses sampled from the learned generative model
    compared with the training mean response.
    (b) Held-out representative IDA responses relative to the generated
    5--95\% response envelope.
    }
    \label{fig:generated_response_validation}
\end{figure}

The generated response ensemble is summarized in Fig.~\ref{fig:generated_response_validation}(a), where gray curves denote generated median IDA responses, the dashed red line denotes their mean, and the black solid line denotes the mean response of the training structural realizations. The close agreement between the two mean curves indicates that the generative model reproduces the central trend of the training response ensemble. Figure~\ref{fig:generated_response_validation}(b) provides an additional held-out validation by comparing the nine validation IDA curves with the unconditional 90\% envelope of the generated response distribution. Across all held-out response ordinates, 92.0\% lie within this envelope.  One held-out response extends partially beyond the envelope, which is not unexpected because the shaded region represents the central 90\% of the unconditional generated distribution rather than a calibrated case-specific prediction interval. Realizations associated with less probable combinations of structural properties may therefore fall outside this range.

\begin{figure}[!htb]
\centering
\includegraphics[width=1.0\textwidth]{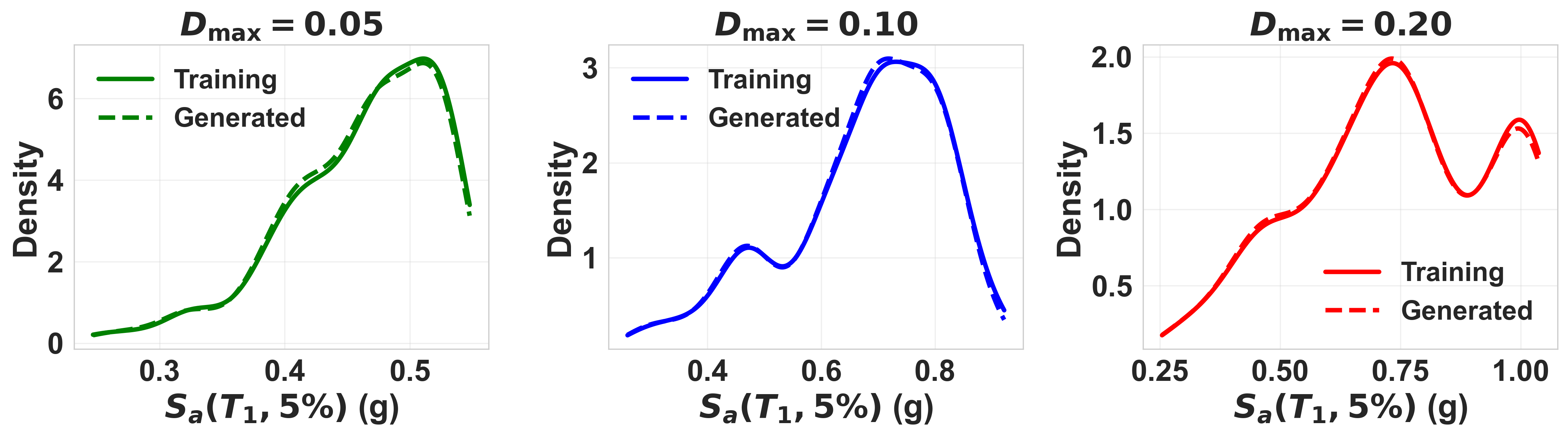}
\caption{Probability densities of the 5\%-damped first-mode spectral acceleration $S_a(T_1,5\%)$ at three values of the maximum interstory drift $D_{\max}$. Solid lines correspond to the training distributions, while dashed lines correspond to the admissible generated samples.}
\label{fig:ida_DMAPS_DENSITIES}
\end{figure}

Fig.~\ref{fig:ida_DMAPS_DENSITIES} compares the probability densities of the 5\%-damped first-mode spectral acceleration $S_a(T_1,5\%)$ at $D_{\max}=0.05$, 0.10, and 0.20. The generated and reference densities remain close across the selected response regimes, including the increasingly non-Gaussian structure observed at larger drift. To assess whether the generated samples preserve the joint dependence between
structural properties and seismic response, we compared the
parameter--response cross-correlation matrices of the training and
generated ensembles. The generated cross-correlation structure closely reproduces that
of the training data, with a relative Frobenius error of $0.0696$, a mean
absolute entry-wise error of $0.0156$, and a correlation coefficient of
$0.9965$ between the vectorized training and generated cross-correlation
matrices. These results indicate that the generative model preserves the
dominant first-order parameter--response dependence in addition to matching
the marginal distributions.

\subsection{Forward Conditional Sampling}
\label{sec:conditional_sampling}

\noindent
An important advantage of learning the joint probability distribution
$f_{\mathbf{x}_s,\mathbf{x}_r}(\cdot)$ is that the resulting generative
representation can be used for conditional inference, rather than being used only
for unconditional generation of synthetic structural-response realizations.
In particular, the joint model enables forward conditional inference of the
form
\begin{equation}
f_{\mathbf{x}_r\mid \mathbf{x}_s}
\left(
\mathbf{x}_r \mid \mathbf{x}_s
\right),
\label{eq:forward_conditional}
\end{equation}
thereby allowing the distribution of seismic response to be evaluated for
prescribed structural properties without requiring additional nonlinear
response-history analyses.

Because conditioning on the complete structural-parameter vector
$\mathbf{x}_s$ would substantially restrict the remaining response
variability, we demonstrate the conditional capability by prescribing a
single physically meaningful structural parameter while allowing the
remaining uncertain parameters to vary. Specifically, we consider the
capping ductility parameter $\mu_c$ and evaluate
\begin{equation}
f_{\mathbf{x}_r\mid \mu_c}
\left(
\mathbf{x}_r\mid \mu_c=c
\right),
\label{eq:conditional_response_muc}
\end{equation}
for representative low, nominal, and high values of $\mu_c$. The selected
conditioning values, $\mu_c = 1.924, 
\mu_c = 3.068,
\mu_c = 4.114,$
correspond to low, nominal, and high regions of the sampled parameter
distribution, respectively.

Conditional statistics are estimated directly from samples generated by the
learned joint model. For a target conditioning value $c$, each generated
realization is assigned a Gaussian kernel weight
\begin{equation}
w_i(c)
=
\exp
\left[
-\frac{1}{2}
\left(
\frac{\mu_{c,i}-c}{h_c}
\right)^2
\right],
\label{eq:conditional_weights_forward}
\end{equation}
where $h_c$ denotes the conditioning bandwidth. In the present analysis, $h_c
=
0.20 \times \mathrm{std}_{\mathrm{train}}(\mu_c)
=
0.2085.$
The normalized weights are then used to estimate the conditional median and
the corresponding response quantiles at each value of maximum interstory drift
$D_{\max}$. The effective sample size associated with each conditional estimate
is computed as
\begin{equation}
N_{\mathrm{eff}}
=
\frac{
\left(
\sum_{i=1}^{N^\star} w_i
\right)^2
}{
\sum_{i=1}^{N^\star} w_i^2
},
\label{eq:effective_sample_size}
\end{equation}
providing a measure of the number of generated samples that contribute
effectively to the conditional estimate.

\begin{figure}[!htb]
    \centering
    \includegraphics[width=1.0\textwidth]{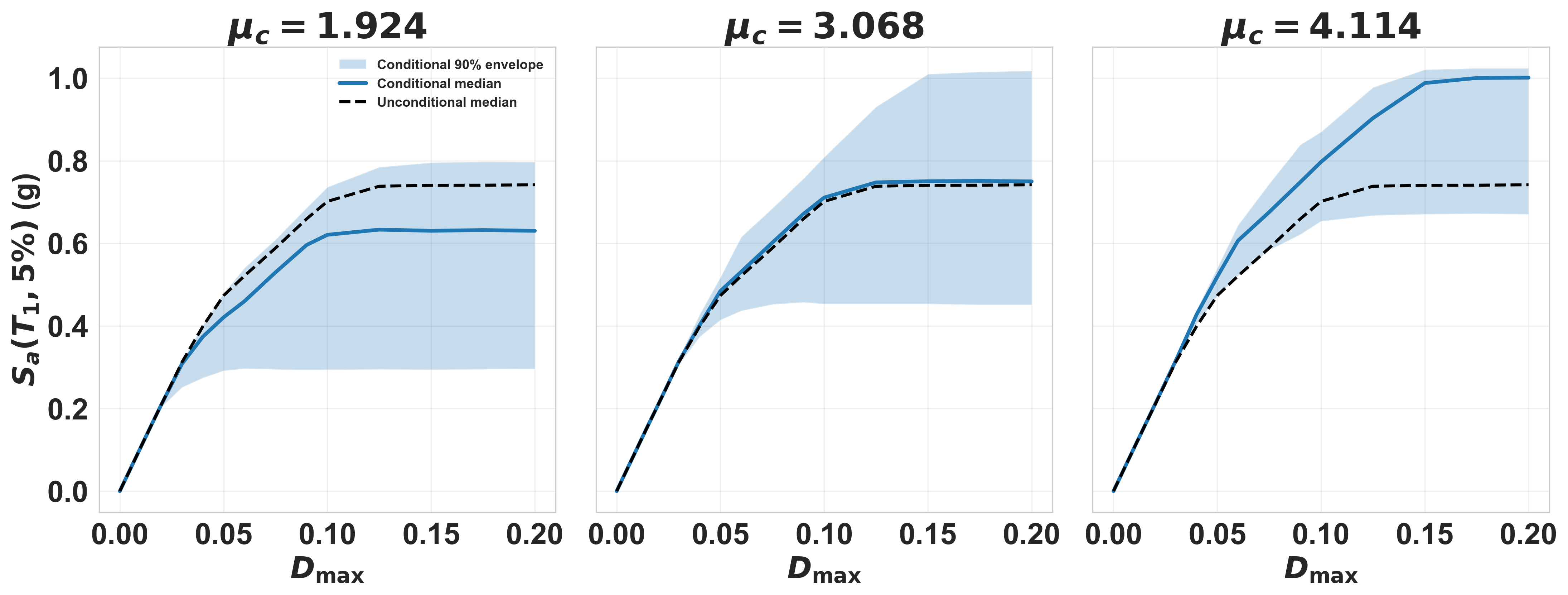}
    \caption{Forward conditional sampling of the representative IDA response for low
    ($\mu_c=1.924$), nominal ($\mu_c=3.068$), and high
    ($\mu_c=4.114$) values of the capping ductility parameter. Solid colored
    lines denote the conditional median responses, shaded regions denote the
    corresponding 5--95\% intervals, and the dashed black line denotes the
    unconditional median response. Conditional statistics are evaluated from
    samples of the learned joint distribution without additional nonlinear
    response-history analyses.}
    \label{fig:conditional_forward}
\end{figure}

Fig.~\ref{fig:conditional_forward} presents the conditional representative IDA
responses together with the corresponding 5--95\% intervals. The
unconditional median response is included as a reference. A clear systematic
dependence on $\mu_c$ is observed. At relatively small drift levels, the
conditional responses remain close to one another, indicating that the
influence of the capping ductility parameter is limited in the predominantly
elastic response regime. As $D_{\max}$ increases and the response enters the
nonlinear range, the conditional distributions separate progressively.
Larger values of $\mu_c$ are associated with higher spectral-acceleration
capacity, whereas smaller values of $\mu_c$ produce lower conditional median
responses. The 5--95\% intervals reflect variability in the learned response
induced by the remaining five uncertain structural parameters; they do not
represent record-to-record ground-motion variability.

These results illustrate a capability that is not directly available from a
conventional deterministic surrogate model. Once the joint distribution
$f_{\mathbf{x}_s,\mathbf{x}_r}(\cdot)$ has been learned, the model can
be queried to characterize response distributions associated with prescribed
structural-property values while accounting for variability in the remaining
uncertain parameters. In this way, the generative representation provides a
computationally efficient mechanism for probabilistic interrogation of the
structural parameter--response relationship across the seismic-performance
range.

\subsection{Inverse Conditional Sampling}
\label{sec:inverse_sampling}

\begin{figure}[!htb]
\centering
\includegraphics[width=0.95\textwidth]{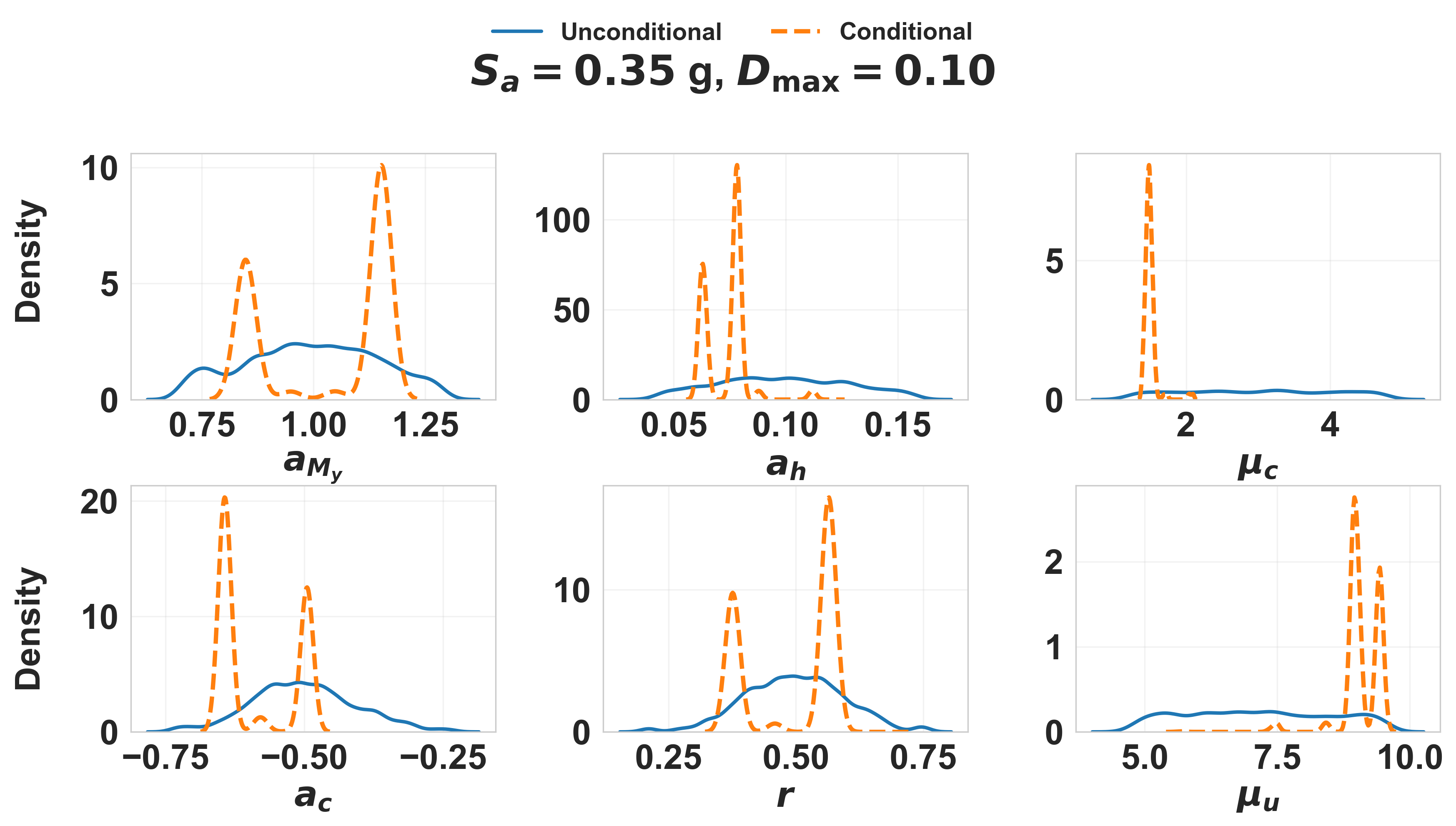}
\caption{Unconditional and response-conditioned structural-parameter distributions for the observation $S_a=0.35$ g at $D_{\max}=0.10$.}
\label{fig:inverse_marginals}
\end{figure}

\noindent
The learned joint distribution can also be inferred in the inverse
direction, by conditioning the structural parameters on an observed response
state. We consider
\begin{equation}
f_{\mathbf{x}_s\mid S_a}
\left(
\mathbf{x}_s
\mid
S_a(D_{\max}=0.10)=0.35\,\mathrm{g}
\right).
\label{eq:inverse_conditional}
\end{equation}
For each generated realization, the Gaussian conditioning weights are
\begin{equation}
w_i
=
\exp
\left[
-\frac{1}{2}
\left(
\frac{
S_{a,i}(0.10)-0.35
}{
h_{S_a}
}
\right)^2
\right],
\label{eq:conditional_weights_inverse}
\end{equation}
with $h_{S_a}
=
0.10\,
\mathrm{std}
\left[
S_a(D_{\max}=0.10)
\right]
=
0.01414\,\mathrm{g}.$
The corresponding effective sample size is $N_{\mathrm{eff}}=138$. Fig.~\ref{fig:inverse_marginals} compares the unconditional structural-parameter
distributions with the response-conditioned distributions. Conditioning on the
single response state produces substantial changes in several parameters,
including strong concentration toward low $\mu_c$ and high $\mu_u$, together
with multimodal structure in a number of the remaining parameters.

The multimodal one-dimensional conditional distributions indicate that the
inverse problem does not admit a unique structural-parameter estimate. A
single response ordinate constrains six uncertain structural variables, and
distinct combinations of these variables may therefore remain compatible
with the same observed performance state. To examine whether the apparent
multimodality reflects coherent structure in the full parameter space, the
response-conditioned samples were further analyzed in the six-dimensional
space $\boldsymbol{\theta}
=
[aM_y,a_h,\mu_c,a_c,r,\mu_u]$.

\begin{figure}[!htb]
\centering
\includegraphics[width=1.0\textwidth]{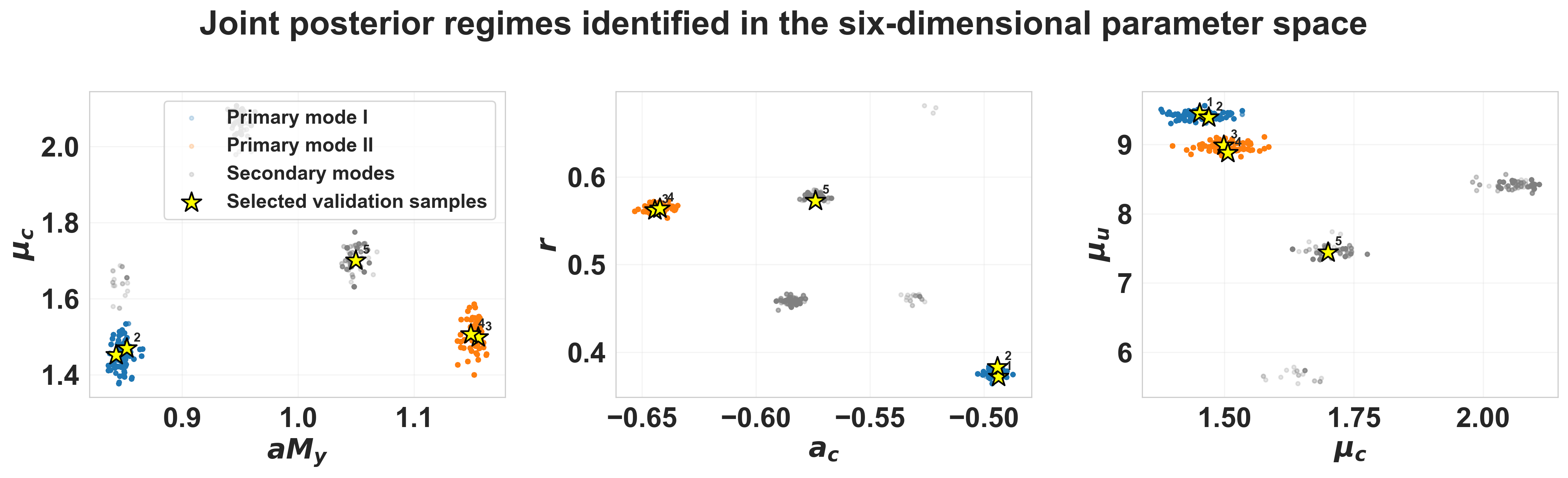}
\caption{Representative projections of the joint response-conditioned
distribution for $S_a=0.35$~g at $D_{\max}=0.10$. Colors denote posterior
regimes identified in the full six-dimensional parameter space. Two dominant
modes contain the majority of the posterior probability mass, while the
remaining disconnected regions correspond to lower-probability secondary
modes.}
\label{fig:inverse_joint}
\end{figure}

The six-dimensional conditional sample cloud was standardized and represented
using a Gaussian mixture model, with the number of mixture components selected
by the Bayesian information criterion (BIC) over the candidate models
considered. Several mixture components occupied closely related regions of
parameter space and were therefore grouped into two dominant posterior
regimes, while the remaining low-weight components were classified as
secondary modes. Together, the two dominant regimes contained approximately
95\% of the conditional probability mass. The resulting joint structure is
illustrated through representative two-dimensional projections in
Fig.~\ref{fig:inverse_joint}. The two dominant regimes remain clearly separated
across these projections, together with several smaller secondary regions, and
correspond to substantially different combinations of $aM_y$, $a_c$, $r$, and
$\mu_u$, while $\mu_c$ remains comparatively similar between them. 

The joint posterior structure can be interpreted more directly by referring to
the post-capping branch of the backbone in Fig.~\ref{fig:side-by-side}. Let
$M_c$ denote the normalized moment at the capping point. For a realization
with parameters $(aM_y,a_h,\mu_c,a_c,r,\mu_u)$,
\begin{equation}
M_c
=
aM_y + a_h(\mu_c-1).
\label{eq:Mc}
\end{equation}
The degrading branch reaches the residual-strength plateau $r$ at a rotation
$\mu_r$ satisfying
\begin{equation}
M_c + a_c(\mu_r-\mu_c)=r,
\end{equation}
and therefore
\begin{equation}
\mu_r
=
\mu_c+\frac{r-M_c}{a_c}.
\label{eq:mur}
\end{equation}
The deformation interval associated with post-capping degradation is then
\begin{equation}
\Delta\mu_{\mathrm{deg}}
=
\mu_r-\mu_c,
\label{eq:delta_mu_deg}
\end{equation}
while the area under the degrading branch is
\begin{equation}
A_{\mathrm{deg}}
=
\frac{M_c+r}{2}
\Delta\mu_{\mathrm{deg}}.
\label{eq:Adeg}
\end{equation}
We also define the area above the residual plateau,
\begin{equation}
A_{\mathrm{excess}}
=
\frac{M_c-r}{2}
\Delta\mu_{\mathrm{deg}},
\label{eq:Aexcess}
\end{equation}
which isolates the additional resistance carried during degradation above the
residual level.

When these quantities are evaluated over the response-conditioned posterior,
the two dominant regimes remain clearly distinct in terms of post-capping
capacity. Primary mode I has a median
$A_{\mathrm{deg}}\approx0.634$, whereas Primary mode II has
$A_{\mathrm{deg}}\approx0.855$. Thus, the same conditioning state,
$(S_a,D_{\max})=(0.35~\mathrm{g},0.10)$, is compatible with two distinct
ranges of integrated post-capping resistance. At the same time, the rotations
at which the degrading branch reaches the residual plateau are nearly
identical, with median values
$\mu_r\approx2.46$ and $\mu_r\approx2.47$ for Primary modes I and II,
respectively. The corresponding degradation intervals are also similar,
$\Delta\mu_{\mathrm{deg}}\approx1.01$ and $0.98$. These results suggest that
the same observed maximum-drift state can arise from different post-capping
strength-capacity regimes while preserving a similar deformation extent over
which degradation develops.

\begin{figure}[!htb]
    \centering
    \includegraphics[width=0.72\textwidth]{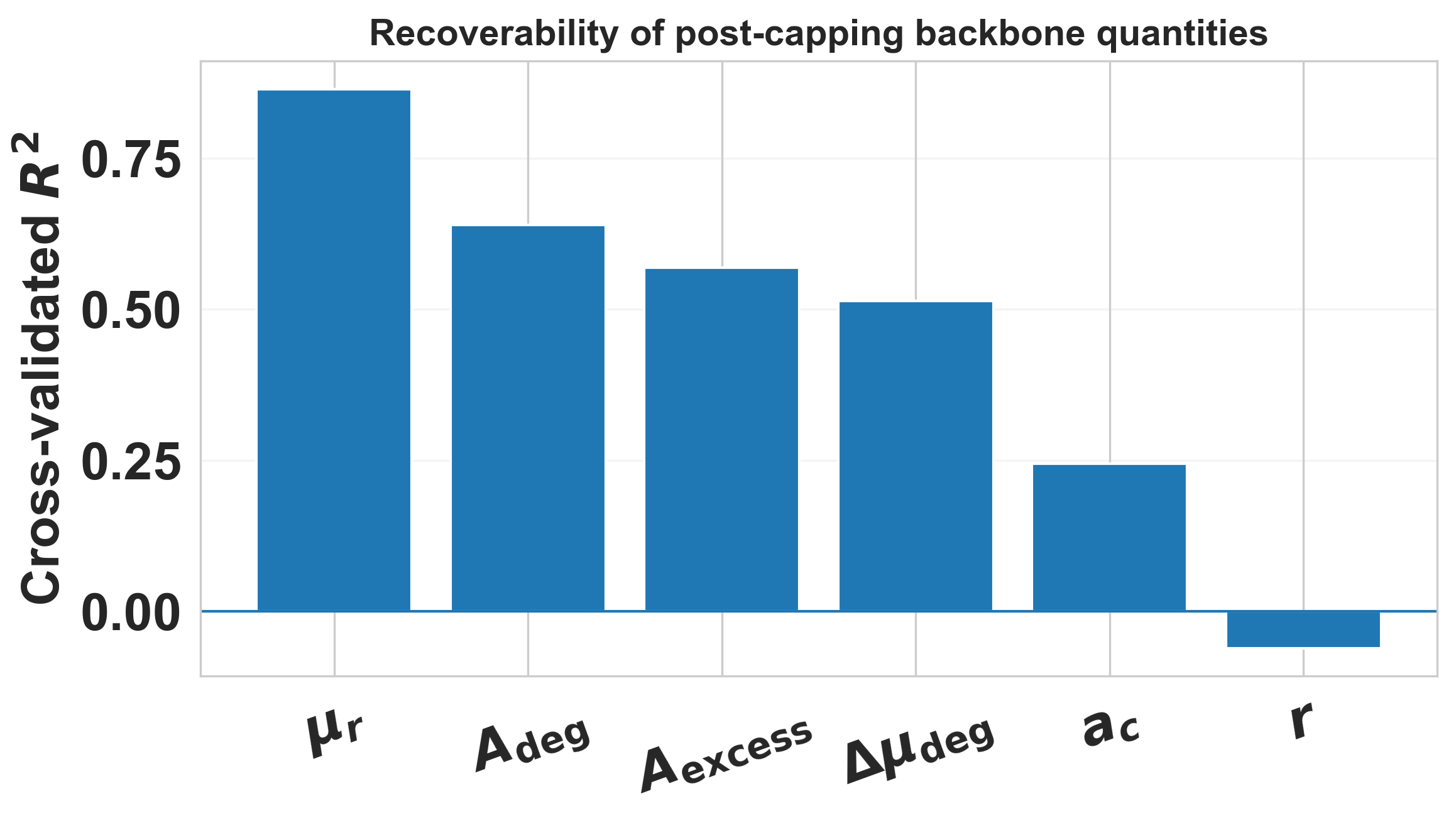}
    \caption{
    Recoverability of individual and derived post-capping backbone quantities
    from the six retained response Diffusion Maps coordinates. The derived
    rotation $\mu_r$, corresponding to the point at which the degrading branch
    reaches the residual-strength plateau, is the most strongly encoded
    quantity. The integrated post-capping measures
    $A_{\mathrm{deg}}$ and $A_{\mathrm{excess}}$ and the degradation interval
    $\Delta\mu_{\mathrm{deg}}$ are also substantially more recoverable than
    the individual parameters $a_c$ and $r$. The reported values are
    out-of-sample coefficients of determination obtained using the same nested
    cross-validation procedure as in Fig.~\ref{fig:physical_manifold}.
    }
    \label{fig:post_capping_recoverability}
\end{figure}

To examine which aspects of this post-capping behavior are actually encoded
in the learned response manifold, the same nested five-fold cross-validation
procedure used for the individual backbone parameters was applied to
$\mu_r$, $\Delta\mu_{\mathrm{deg}}$, $A_{\mathrm{deg}}$, and
$A_{\mathrm{excess}}$. The results are summarized in
Fig.~\ref{fig:post_capping_recoverability}. The rotation $\mu_r$ is by far
the most recoverable quantity, with $R^2\approx0.865$. The integrated
post-capping quantities are also strongly recoverable, with
$R^2\approx0.640$ for $A_{\mathrm{deg}}$ and
$R^2\approx0.569$ for $A_{\mathrm{excess}}$, while
$\Delta\mu_{\mathrm{deg}}$ gives $R^2\approx0.513$. By comparison, the
individual post-capping slope $a_c$ has only modest recoverability
($R^2\approx0.245$), and the residual-strength parameter $r$ is essentially
unrecoverable on its own ($R^2\approx-0.061$).

Taken together, these results indicate that the response manifold is not
primarily encoding the residual-strength level $r$ as an isolated parameter.
Instead, the maximum-drift IDA response is substantially more informative
about the location and evolution of post-capping degradation and about
integrated measures of the resistance retained during this process. In
particular, the strong recoverability of $\mu_r$ suggests that the observable
response is sensitive to when the degrading branch reaches the residual
regime, whereas the precise height of the idealized residual plateau is only
weakly identifiable. This also explains how $r$ can participate in the joint
posterior structure despite being individually unrecoverable: different
combinations of degradation slope and residual strength can produce distinct
post-capping capacity profiles while remaining compatible with the same
global response state.

This
observation is consistent with the stronger recoverability of $\mu_c$ found
previously and suggests that the observable IDA response constrains some
parameter directions more strongly than others. The joint posterior structure also helps reconcile the parameter-recoverability
results of Fig.~\ref{fig:physical_manifold} with the multimodal inverse
distributions. In particular, the bimodality observed in $a_c$ is accompanied
by a systematic shift in the residual-strength parameter $r$ between the two
dominant posterior regimes. Thus, different combinations of post-capping
slope and residual strength can produce comparable response at the
conditioning state. This does not contradict the low individual recoverability
of $r$ in Fig.~\ref{fig:physical_manifold}. The recoverability analysis asks
whether $r$ can be inferred uniquely from the response manifold, whereas the
joint inverse analysis reveals that $r$ participates in compensating parameter
combinations. Hence, $r$ may be weakly identifiable individually while still
playing an important role in the joint conditional structure.

Referring to the backbone response in Fig.~\ref{fig:side-by-side}(b),
$a_c$ controls the rate of post-capping strength degradation, whereas $r$
defines the residual-strength plateau reached after substantial degradation.
The stronger recoverability of $a_c$ therefore suggests that the maximum-drift
IDA response is more informative about the onset and progression of
post-capping degradation than about the precise residual strength attained at
later stages. This is physically plausible because $r$ represents an
idealized description of severely degraded behavior and may contribute
relatively little additional information to the global maximum interstory drift
response. At the same time, the joint posterior indicates that changes in
$a_c$ are accompanied by systematic changes in $r$, so that different
combinations of degradation rate and residual strength can produce similar
observable IDA response. This provides a physical interpretation of the
inverse non-uniqueness: $r$ is weakly identifiable on its own, but variations
in residual strength can compensate for changes in the better-constrained
post-capping slope.

\begin{figure}[!htb]
\centering
\includegraphics[width=0.8\textwidth]{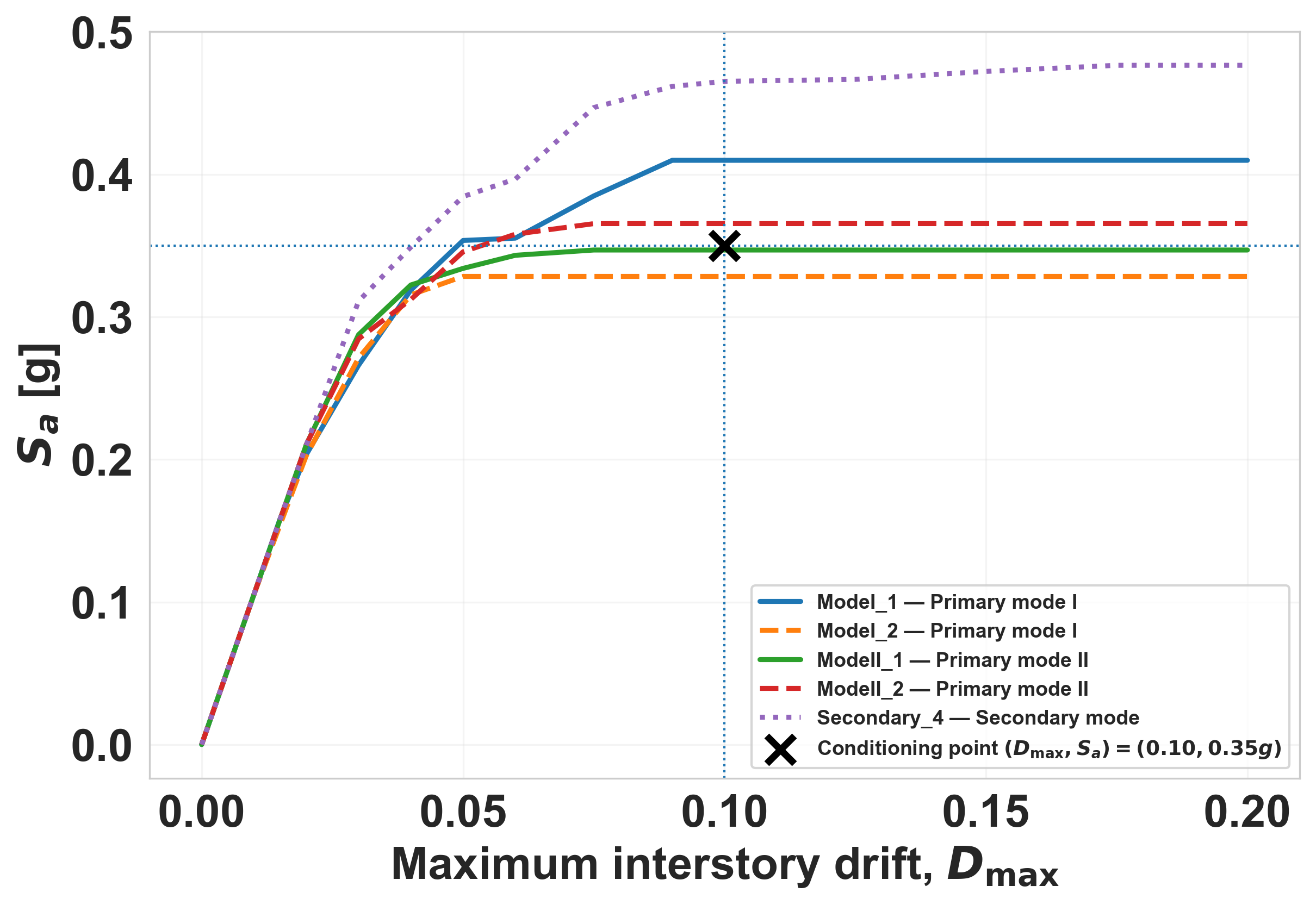}
\caption{Forward-model validation of representative parameter realizations
selected from the response-conditioned posterior at
$S_a=0.35$~g and $D_{\max}=0.10$. Each curve is the median IDA response over
the same 30 ground-motion records used in the original dataset. The two
dominant posterior modes contain realizations that reproduce the conditioning
response with varying accuracy, whereas the tested secondary low-probability
mode exhibits a larger deviation. The black marker denotes the conditioning
point.}
\label{fig:ida_valid}
\end{figure}

The separation of the posterior modes alone does not establish that all of
these regions correspond to physically admissible structural states, since
some disconnected branches could arise from the generative approximation.
To assess this directly, five samples were selected from the
conditional distribution: two representative realizations from each dominant
mode and one realization from a secondary low-probability mode. Each
parameter vector was then propagated independently through the original
OpenSees model using the same 30-record adaptive IDA procedure employed to
construct the training dataset. The resulting IDA curves were postprocessed
using the same common-drift interpolation and record-wise median procedure as
the original data.
Figure~\ref{fig:ida_valid} compares the resulting median IDA curves with the
conditioning point $(D_{\max},S_a)=(0.10,0.35~\mathrm{g})$. The two
realizations drawn from Primary mode II give
$S_a(D_{\max}=0.10)=0.347$ and $0.366$~g, corresponding to relative errors of
approximately $0.85\%$ and $4.44\%$, respectively. The two realizations from
Primary mode I give $0.410$ and $0.329$~g, showing larger within-mode
variability but remaining in the vicinity of the conditioning response. In
contrast, the representative secondary-mode realization gives
$S_a(D_{\max}=0.10)=0.465$~g, approximately $33\%$ above the conditioning
value. These forward-model evaluations provide evidence that the two dominant
disconnected regions of the conditional distribution contain distinct
structural parameter combinations capable of producing comparable observable
IDA behavior. In this sense, the multimodality is consistent with
non-uniqueness of the inverse mapping: substantially different parameter
vectors can correspond to similar response states. The poorer agreement of
the tested secondary mode further suggests that not all low-probability
branches are equally consistent with the original forward model. Because only
a small number of representative realizations are examined here, these
results are interpreted as a validation of the existence of alternative
admissible parameter regimes.

The inverse results were also examined for sensitivity to the conditioning
bandwidth over the range
$h_{S_a}/\sigma_{S_a}=0.075$--0.20. The dominant concentration toward low
$\mu_c$ and high $\mu_u$ remains, while some marginal modes and quantiles
broaden as the bandwidth increases. Taken together with the multimodal joint
structure and the forward-model validation above, these results support
interpreting the inverse solution as a response-conditioned probability
distribution over admissible structural states rather than as a unique
parameter estimate.

\section{Discussion}
\label{sec:discussion}

The results demonstrate that manifold-based generative
learning can complement high-fidelity nonlinear
earthquake simulations by providing a probabilistic
representation of the relationship between uncertain
structural properties and seismic performance.
Rather than approximating a deterministic mapping
from structural parameters to selected response quantities,
the proposed framework learns their joint probability
distribution. This distinction enables the generation
of additional parameter--response realizations and
supports probabilistic inference in both directions:
predicting seismic response for prescribed structural
properties and inferring plausible structural configurations
from observed performance. By combining nonlinear manifold learning,
score-based diffusion modeling, and Geometric Harmonics,
these capabilities are obtained from a limited ensemble
of nonlinear structural simulations, without requiring
additional nonlinear analyses during generation.

The numerical results for the nine-story steel
moment-resisting frame demonstrate that the learned
representation reproduces important characteristics
of the original simulation ensemble. The six-dimensional
response manifold reconstructs the nine held-out
representative IDA curves with an overall root-mean-square
error of $0.0106$~g. Furthermore, $92.0\%$ of the
held-out responses fall within the unconditional
5--95\% envelope of the generated response distribution.
The generative model also reproduces the marginal
distributions of the structural parameters, including
the lower tail of the capping ductility parameter
$\mu_c$, and closely preserves the parameter--response
cross-correlation structure. These results indicate
that the learned model captures important features
of both the variability in seismic performance and
its dependence on uncertain structural properties.
Nevertheless, agreement in marginal distributions
and low-order dependence measures does not guarantee
that all aspects of the joint distribution are accurately
represented, particularly in regions of parameter
space sparsely covered by the training data.

An important finding concerns the information about
structural properties that can be recovered from
the representative IDA responses. Although the
response manifold provides an accurate low-dimensional
description of seismic performance, its coordinates
do not encode all structural parameters equally.
The capping ductility parameter $\mu_c$ is the most
recoverable individual parameter, whereas the
post-capping slope $a_c$ is only modestly recoverable
and the residual-strength parameter $r$ is essentially
unrecoverable from the selected response representation.
These differences should not be interpreted as
conventional forward sensitivities of the structural
model. Rather, they characterize how distinctly
variations in the structural parameters are expressed
in the observable IDA responses. More importantly, the results show that physically
meaningful combinations of structural parameters
can be substantially more recoverable than the
individual parameters themselves. The rotation
$\mu_r$ at which the degrading backbone reaches
the residual-strength plateau achieves an
out-of-sample coefficient of determination of
$R^2\approx0.865$. The degradation interval
$\Delta\mu_{\mathrm{deg}}$ and the integrated
post-capping resistance measures
$A_{\mathrm{deg}}$ and $A_{\mathrm{excess}}$
are also more strongly recoverable than the
individual post-capping slope and residual-strength
parameters. This suggests that the representative
IDA relation $S_a(D_{\max})$ contains more information
about the location, extent, and integrated resistance
characteristics of post-capping degradation than
about the precise values of the parameters defining
the degrading branch. Consequently, poor
recoverability of an individual parameter does
not necessarily imply that the associated physical
mechanism has little influence on seismic performance.
Instead, the observable response may constrain
particular combinations of parameters while
remaining relatively insensitive to their
individual values.

This distinction is central to the interpretation
of structural identifiability. From a mathematical
perspective, identifiability concerns whether
distinct parameter configurations produce
distinguishable model responses. The present
recoverability analysis does not establish
formal structural identifiability, but it
provides evidence that the information available
in the selected seismic-performance representation
is distributed unevenly among the physical
parameters and their combinations. In particular,
the greater recoverability of derived post-capping
quantities suggests that alternative parameterizations
based on observable characteristics of the degrading
backbone may be more informative for structural
model updating than the original backbone
parameters considered individually.

The conditional inference results further illustrate
the implications of learning the joint distribution.
Forward conditioning on $\mu_c$ produces systematic
changes in the representative IDA response.
At relatively small drift levels, the conditional
responses remain close, whereas differences
become more pronounced as the structure enters
the nonlinear regime. Larger capping ductility
is associated with increased spectral-acceleration
capacity at large drift, consistent with the
physical role of $\mu_c$ in delaying the onset
of post-capping degradation. Thus, the learned
joint distribution captures physically meaningful
relationships between structural properties
and seismic performance while accounting for
variability in the remaining uncertain parameters.

Inverse conditioning provides a complementary
perspective by revealing the ambiguity of
structural parameter inference from observed
performance. For the conditioning state
$(S_a,D_{\max})=(0.35~\mathrm{g},0.10)$,
the inferred parameter distribution contains
two dominant, separated regions in the
six-dimensional structural-parameter space.
These regions correspond to substantially
different combinations of post-capping slope
and residual strength, yet both remain
compatible with the prescribed response state.
Interestingly, they exhibit different levels
of integrated post-capping resistance while
reaching the residual-strength plateau at
nearly the same rotation and over similar
degradation intervals. This demonstrates
that a single IDA performance state may
constrain certain characteristics of
post-capping behavior without uniquely
determining the underlying strength-degradation
mechanism.

To further assess the inferred parameter configurations, representative samples from the conditional distribution were evaluated using the numerical model.
Samples from both dominant modes produced IDA responses close to the prescribed conditioning state, although the agreement varied among realizations. In contrast, the sample from the secondary mode exhibited a larger deviation from the target response. These results provide
independent evidence that different combinations of structural parameters can produce similar seismic performance, while also revealing discrepancies between the learned distribution and the original nonlinear model.
The latter highlights the need for physics-based validation of generated parameter configurations, particularly when interpreting low-probability regions of the
conditional distribution.

The choice of generative representation
also has important methodological implications.
By applying Diffusion Maps only to the seismic
responses, the framework reduces the dimensionality
of the response trajectories while retaining
the six structural parameters explicitly.
Consequently, the structural parameters can
be sampled and conditioned upon directly,
without requiring their reconstruction from
the learned response manifold. This distinction
is particularly relevant when the mapping
from structural properties to observable
performance is non-unique. Nevertheless,
the reliability of the generated responses
remains dependent on the accuracy and support
of the learned manifold and its reconstruction
through Geometric Harmonics.

Several limitations should be considered
when interpreting these findings. First,
the analysis is restricted to a single
nine-story steel moment-resisting frame
and six uncertain nonlinear backbone
parameters. Although this benchmark provides
a physically meaningful setting for evaluating
the methodology, the findings cannot
automatically be generalized to other
structural systems, modeling assumptions,
or earthquake environments. Second, the
generative model is learned from a relatively
small dataset, which may limit the accuracy
of the estimated joint distribution in
sparsely sampled regions. Third, the
inverse conditional distributions depend
on the conditioning bandwidth, and the
independent forward-model validation of
the identified conditional regimes is
based on a limited number of representative
parameter configurations. More extensive
validation is therefore required to establish
the reliability of the inferred distributions
throughout parameter space. An additional limitation concerns the
representation of ground-motion uncertainty.
Although each representative IDA response
is constructed from 30 ground-motion records,
record-to-record variability is aggregated
through the pointwise median rather than
represented explicitly in the learned
generative state. The resulting model
therefore characterizes uncertainty in
structural properties and their representative
seismic-performance responses, but does
not provide a complete probabilistic
description of both structural uncertainty
and ground-motion variability. In particular,
the conditional response intervals should
not be interpreted as capturing the full
variability of individual earthquake
response histories.

Future developments should extend the joint generative
representation to include ground-motion descriptors and
record-level responses, allowing a more complete
characterization of the interaction between structural
uncertainty and earthquake excitation. A further
direction is the systematic investigation of structural
parameter identifiability using complementary analytical
and data-driven approaches. In particular, Jacobian-based
sensitivity analysis and the associated Fisher information
matrix could help identify poorly constrained parameter
directions and establish how the identifiable combinations
inferred from the response manifold relate to the
underlying structural model. Incorporating additional
engineering demand measures and response observations
could further improve parameter recoverability and help
distinguish structural configurations that produce similar
maximum-interstory-drift responses. Finally, embedding
physical constraints into the generative process could
reduce inadmissible samples, while validation across
different structural typologies and larger independent
datasets would help establish the generalizability and
reliability of the proposed framework.

Overall, this study demonstrates that
manifold-based generative learning can
extend conventional seismic surrogate
modeling from forward response prediction
to probabilistic characterization of
structural behavior and inverse inference.
Beyond reproducing the distribution of
seismic responses, the framework reveals
which aspects of nonlinear structural
behavior are recoverable from the selected
performance observables and which remain
ambiguous. These capabilities provide
a basis for uncertainty-aware structural
model updating and seismic performance
assessment, while emphasizing that
generative predictions must remain
grounded in the information content
of the available observations and
validated against the underlying
physics-based simulations.

\appendix

\section{Diffusion Maps}
\label{sec:DMaps}

\subsection{Diffusion Maps for Dimensionality Reduction}

\noindent
Diffusion Maps \cite{coifman2006diffusion} is a nonlinear dimensionality reduction technique that captures the intrinsic geometry of a dataset \( \mathcal{X} = \{\mathbf{x}_1, \mathbf{x}_2, \ldots, \mathbf{x}_N\} \) by constructing a local similarity graph among the data points. A kernel-based affinity matrix \( \mathbf{K} \in \mathbb{R}^{N \times N} \) is defined using the following diffusion kernel:
\begin{equation}
    K(\mathbf{x}_i, \mathbf{x}_j) = \exp\left(-\frac{\| \mathbf{x}_i - \mathbf{x}_j \|^2}{2\epsilon}\right),
\end{equation}
where \( \epsilon \) is a kernel bandwidth parameter \cite{gear2016manifolds}. Data points \( \mathbf{x}_i \) and \( \mathbf{x}_j \) are considered connected if their squared distance is less than \( \epsilon \). To correct for sampling density and recover the Laplace-Beltrami operator \( \Delta \) on the underlying manifold \( \mathcal{M} \), normalization is applied:
\begin{equation}
    \tilde{\mathbf{K}} = \mathbf{P}^{-\alpha} \mathbf{K} \mathbf{P}^{-\alpha},
\end{equation}
where \( \mathbf{P} \) is a diagonal matrix with \( P_{ii} = \sum_{j=1}^N K_{ij} \), and \( \alpha \) adjusts the influence of sampling density. A further normalization produces a row-stochastic matrix:
\begin{equation}
    \mathbf{M} = \mathbf{D}^{-1} \tilde{\mathbf{K}}, \quad D_{ii} = \sum_{j=1}^N \tilde{K}_{ij}.
\end{equation}
The eigendecomposition of \( \mathbf{M} \) yields eigenvectors \( \boldsymbol{\phi}_i \) and eigenvalues \( \lambda_i \) via:
\begin{equation}
    \mathbf{M} \boldsymbol{\phi}_i = \lambda_i \boldsymbol{\phi}_i.
\end{equation}

A reduced representation is constructed from non-harmonic eigenvectors, excluding those that are harmonic or redundant. These selected eigenvectors \( \boldsymbol{\Phi} = \{\boldsymbol{\phi}_{i_1}, \ldots, \boldsymbol{\phi}_{i_k}\} \) provide a parsimonious parametrization of the data when \( k \ll d \).

To assess the sensitivity of the response representation to the Diffusion Maps
kernel scale, we considered
\[
\epsilon_r
=
c\times\mathrm{median}_{i<j}
\left\|
\mathbf{x}_{r,i}^{(\mathrm{std})}
-
\mathbf{x}_{r,j}^{(\mathrm{std})}
\right\|^2,
\qquad
c\in\{0.5,1,2,5,10\}.
\]
For each value of $c$, six non-harmonic coordinates were retained and the
response was reconstructed using Geometric Harmonics. The corresponding
held-out reconstruction RMSEs were $0.0254$, $0.0196$, $0.0235$, $0.0159$,
and $0.0106$~g, respectively. Thus, the reconstruction quality improves for
the broader kernels considered here, with the best performance obtained for
$c=10$. Because the selected value corresponds to a kernel
($K_{ij}\approx0.95$ at the median pairwise distance), we also compared the
6-dimensional Diffusion Maps representation with a 6-component PCA
baseline. PCA yielded a held-out reconstruction RMSE of $0.0104$~g, which is
comparable to the $0.0106$~g obtained with Diffusion Maps. The present results
therefore do not indicate a clear reconstruction advantage of the nonlinear
embedding over PCA. Rather, Diffusion Maps provides a compact nonlinear
coordinate system that is used within the manifold-based generative framework,
while the subsequent generative ablation evaluates whether this representation
provides advantages beyond reconstruction alone.

\subsection{Diffusion Maps for Function Lifting (Double Diffusion Maps)}

\noindent
The Double Diffusion Maps method \cite{evangelou2023double} extends the use of Diffusion Maps to both dimensionality reduction and function extension. Starting from a reduced set of non-harmonic coordinates \( \boldsymbol{\Phi} \), a new basis \( \boldsymbol{\Psi} \), known as \textit{Latent Harmonics}, is constructed by applying a second Diffusion Maps algorithm:
\begin{equation}
    K^*(\boldsymbol{\phi}_i, \boldsymbol{\phi}_j) = \exp\left(-\frac{\|\boldsymbol{\phi}_i - \boldsymbol{\phi}_j\|^2}{2\epsilon_2}\right).
\end{equation}
The eigenvectors \( \boldsymbol{\Psi} = \{\boldsymbol{\psi}_0, \ldots, \boldsymbol{\psi}_{m-1}\} \) and eigenvalues \( \boldsymbol{\sigma} \) are computed. A function \( h \) defined on the dataset is projected as:
\begin{equation}
    P_\delta h = \sum_{j=1}^m \langle h, \boldsymbol{\psi}_j \rangle \boldsymbol{\psi}_j.
\end{equation}
Latent Harmonics at a new point \( \boldsymbol{\phi}_\text{new} \) are evaluated via:
\begin{equation}
    \boldsymbol{\Psi}_j(\boldsymbol{\phi}_\text{new}) = \sigma_j^{-1} \sum_{i=1}^m K^*(\boldsymbol{\phi}_\text{new}, \boldsymbol{\phi}_i) \psi_j(\boldsymbol{\phi}_i).
\end{equation}
The extended function value at \( \boldsymbol{\phi}_\text{new} \) is then estimated as:
\begin{equation}
    (Ef)(\boldsymbol{\phi}_\text{new}) = \sum_{j=1}^m \langle h, \boldsymbol{\psi}_j \rangle \boldsymbol{\Psi}_j(\boldsymbol{\phi}_\text{new}).
\end{equation}

\begin{algorithm}[H]
\caption{Double Diffusion Maps}
\label{alg:dim_reduction}
\begin{algorithmic}[1]
\Require Dataset \( \mathcal{X} = \{\mathbf{x}_i\}_{i=1}^N \subset \mathbb{R}^d \sim f_{\mathbf{x}}(\cdot) \)

\State \textbf{Step 1: Dimensionality Reduction with Diffusion Maps}
\begin{itemize}
    \item Compute kernel \( K(\mathbf{x}_i, \mathbf{x}_j) \)
    \item Normalize: \( \tilde{\mathbf{K}} = \mathbf{P}^{-\alpha} \mathbf{K} \mathbf{P}^{-\alpha} \), with \( \alpha = 1 \)
    \item Row-normalize: \( \mathbf{M} = \mathbf{D}^{-1} \tilde{\mathbf{K}} \)
    \item Eigendecompose \( \mathbf{M} \) and select \( \boldsymbol{\Phi} \in \mathbb{R}^{N \times k} \)
\end{itemize}

\State \textbf{Step 2: Latent Harmonics via Double Diffusion Maps}
\begin{itemize}
    \item Construct kernel \( K^*(\boldsymbol{\phi}_i, \boldsymbol{\phi}_j) \)
    \item Compute top \( l \) eigenvectors \( \boldsymbol{\Psi} \in \mathbb{R}^{N \times l} \)
    \item Project \( f \rightarrow P_\delta f \)
\end{itemize}

\State \textbf{Step 3: Function Extension}
\begin{itemize}
    \item Evaluate \( \boldsymbol{\Psi}_j(\boldsymbol{\phi}_\text{new}) \)
    \item Compute \( (Ef)(\boldsymbol{\phi}_\text{new}) \)
\end{itemize}
\end{algorithmic}
\end{algorithm}

\section{Geometric Harmonics for Function Extension}
\label{sec:GH}

\noindent
Geometric Harmonics (GH) extends a function \( h \) defined on a dataset \( \mathcal{X} \) to new points \( \mathbf{x}_\text{new} \notin \mathcal{X} \) using a kernel-based interpolation inspired by the Nystr\"om method \cite{fowlkes2001efficient}. The mapping \( \mathbf{x}_\text{new} \mapsto \boldsymbol{\phi}_\text{new} \) (restriction) is computed as:
\begin{align}
    \mathbf{K}(\mathbf{x}_\text{new}, \mathbf{x}_j) &= \exp\left(-\frac{\|\mathbf{x}_\text{new} - \mathbf{x}_j\|^2}{2\epsilon}\right), \\
    \tilde{\mathbf{K}}(\mathbf{x}_\text{new}, \mathbf{x}_j) &= \frac{\mathbf{K}(\mathbf{x}_\text{new}, \mathbf{x}_j)}{p(\mathbf{x}_\text{new})^\alpha p(\mathbf{x}_j)^\alpha},
\end{align}
where \( p(\mathbf{x}_\text{new}) = \sum_j \mathbf{K}(\mathbf{x}_\text{new}, \mathbf{x}_j) \) and \( p(\mathbf{x}_j) = \sum_k \mathbf{K}(\mathbf{x}_j, \mathbf{x}_k) \). The interpolation weights are defined as:
\begin{equation}
    \mathbf{A}(\mathbf{x}_\text{new}, \mathbf{x}_i) = \frac{\tilde{\mathbf{K}}(\mathbf{x}_\text{new}, \mathbf{x}_i)}{\sum_{j=1}^N \tilde{\mathbf{K}}(\mathbf{x}_\text{new}, \mathbf{x}_j)}.
\end{equation}
The extended eigenfunction value for component \( \beta \) is given by:
\begin{equation}
    \phi_\beta(\mathbf{x}_\text{new}) = \frac{1}{\lambda_\beta} \sum_{i=1}^N \mathbf{A}(\mathbf{x}_\text{new}, \mathbf{x}_i) \phi_\beta(\mathbf{x}_i).
\end{equation}

\bibliographystyle{elsarticle-num}
\bibliography{manuscript}

@article{coifman2006diffusion,
  title={Diffusion maps},
  author={Coifman, RR and Lafon, S},
  journal={\text{Applied and Computational Harmonic Analysis}},
  volume={21},
  number={1},
  pages={5--30},
  year={2006},
  publisher={Elsevier}
}

@article{rothenberg1971identification,
  author  = {Rothenberg, Thomas J.},
  title   = {Identification in Parametric Models},
  journal = {Econometrica},
  volume  = {39},
  number  = {3},
  pages   = {577--591},
  year    = {1971},
  doi     = {10.2307/1913267}
}

@misc{brouwer2018underlying,
  author        = {Brouwer, Andrew F. and Eisenberg, Marisa C.},
  title         = {The Underlying Connections Between Identifiability,
                   Active Subspaces, and Parameter Space Dimension Reduction},
  year          = {2018},
  eprint        = {1802.05641},
  archivePrefix = {arXiv},
  primaryClass  = {math.DS},
  doi           = {10.48550/arXiv.1802.05641},
  url           = {https://arxiv.org/abs/1802.05641}
}

@article{fragiadakis2006evaluation,
  author  = {Fragiadakis, M. and Vamvatsikos, D. and Papadrakakis, M.},
  title   = {Evaluation of the Influence of Vertical Irregularities on the Seismic Performance of a 9-Storey Steel Frame},
  journal = {Earthquake Engineering \& Structural Dynamics},
  year    = {2006},
  volume  = {35},
  number  = {12},
  pages   = {1489--1509}
}

@article{song2020score,
  title={Score-based generative modeling through stochastic differential equations},
  author={Song, Yang and Sohl-Dickstein, Jascha and Kingma, Diederik P and Kumar, Abhishek and Ermon, Stefano and Poole, Ben},
  journal={arXiv preprint arXiv:2011.13456},
  year={2020}
}

@article{vincent2011connection,
  title={A connection between score matching and denoising autoencoders},
  author={Vincent, Pascal},
  journal={Neural computation},
  volume={23},
  number={7},
  pages={1661--1674},
  year={2011},
  publisher={MIT Press}
}

@article{li2019probabilistic,
  title={Probabilistic performance-based optimum seismic design framework: illustration and validation},
  author={Li, Yong and Conte, Joel P and Gill, Philip E},
  journal={Computer Modeling in Engineering \& Sciences},
  volume={120},
  number={3},
  pages={517--543},
  year={2019},
  publisher={Tech Science Press}
}

@article{stojadinovici2018case,
  title={The case for systemic resilience: Urban communities in natural disasters},
  author={Stojadinovi{\'c}i, Bo{\v{z}}idar},
  journal={Domains of resilience for complex interconnected systems.},
  pages={141},
  year={2018}
}

@article{giovanis2025accelerating,
  title={Accelerating uncertainty quantification in incremental dynamic analysis using dimension reduction-based surrogate modeling},
  author={Giovanis, Dimitris G and Taflanidis, Alexandros and Shields, Michael D},
  journal={Bulletin of Earthquake Engineering},
  volume={23},
  number={1},
  pages={391--410},
  year={2025},
  publisher={Springer}
}

@article{giovanis2025generative,
  title   = {Generative learning of densities on manifolds},
  author  = {Giovanis, Dimitris G. and Crabtree, Ellis and Ghanem, Roger G. and Kevrekidis, Ioannis G.},
  journal = {Computer Methods in Applied Mechanics and Engineering},
  volume  = {446},
  pages   = {118266},
  year    = {2025},
  doi     = {10.1016/j.cma.2025.118266},
  publisher = {Elsevier}
}

@article{giovanis2025enabling,
  title   = {Enabling probabilistic learning on manifolds through double diffusion maps},
  author  = {Giovanis, Dimitris G. and Evangelou, Nikolaos and Kevrekidis, Ioannis G. and Ghanem, Roger G.},
  journal = {Journal of Computational Physics},
  volume  = {551},
  pages   = {114663},
  year    = {2026},
  doi     = {10.1016/j.jcp.2026.114663},
  publisher = {Elsevier}
}

@article{liu2024diffusion,
  title={Diffusion-Model-Assisted Supervised Learning of Generative Models for Density Estimation},
  author={Liu, Yanfang and Yang, Minglei and Zhang, Zezhong and Bao, Feng and Cao, Yanzhao and Zhang, Guannan},
  journal={Journal of Machine Learning for Modeling and Computing},
  volume={5},
  number={1},
  year={2024},
  publisher={Begel House Inc.}
}

@article{evangelou2023double,
  title={Double diffusion maps and their latent harmonics for scientific computations in latent space},
  author={Evangelou, Nikolaos and Dietrich, Felix and Chiavazzo, Eliodoro and Lehmberg, Daniel and Meila, Marina and Kevrekidis, Ioannis G},
  journal={Journal of Computational Physics},
  volume={485},
  pages={112072},
  year={2023},
  publisher={Elsevier}
}

@inproceedings{gear2016manifolds,
  title={Manifolds defined by points: Parameterizing and boundary detection},
  author={Gear, CW and Chiavazzo, Eliodoro and Kevrekidis, Yannis G},
  booktitle={AIP Conference Proceedings},
  year={2016}
}

@inproceedings{fowlkes2001efficient,
  title={Efficient spatiotemporal grouping using the nystrom method},
  author={Fowlkes, Charless and Belongie, Serge and Malik, Jitendra},
  booktitle={Proceedings of the 2001 IEEE Computer Society Conference on Computer Vision and Pattern Recognition. CVPR 2001},
  volume={1},
  pages={I--I},
  year={2001},
  organization={IEEE}
}

@article{soize2022probabilistic,
  title={Probabilistic learning on manifolds (PLoM) with partition},
  author={Soize, Christian and Ghanem, Roger},
  journal={International Journal for Numerical Methods in Engineering},
  volume={123},
  number={1},
  pages={268--290},
  year={2022},
  publisher={Wiley Online Library}
}

@article{song2019generative,
  title={Generative modeling by estimating gradients of the data distribution},
  author={Song, Yang and Ermon, Stefano},
  journal={Advances in neural information processing systems},
  volume={32},
  year={2019}
}

@article{ho2020denoising,
  title={Denoising diffusion probabilistic models},
  author={Ho, Jonathan and Jain, Ajay and Abbeel, Pieter},
  journal={Advances in neural information processing systems},
  volume={33},
  pages={6840--6851},
  year={2020}
}

@article{song2020improved,
  title={Improved techniques for training score-based generative models},
  author={Song, Yang and Ermon, Stefano},
  journal={Advances in neural information processing systems},
  volume={33},
  pages={12438--12448},
  year={2020}
}

@misc{songdiffmodels,
  doi = {10.48550/ARXIV.2011.13456},
  
  url = {https://arxiv.org/abs/2011.13456},
  
  author = {Song, Yang and Sohl-Dickstein, Jascha and Kingma, Diederik P. and Kumar, Abhishek and Ermon, Stefano and Poole, Ben},
  
  title = {Score-Based Generative Modeling through Stochastic Differential Equations},
  
  publisher = {arXiv},
  
  year = {2020},
  
  copyright = {arXiv.org perpetual, non-exclusive license}
}

@article{vamvatsikos2014seismic,
  title={Seismic performance uncertainty estimation via IDA with progressive accelerogram-wise latin hypercube sampling},
  author={Vamvatsikos, D},
  journal={Journal of Structural Engineering},
  volume={140},
  number={8},
  pages={A4014015},
  year={2014},
  publisher={American Society of Civil Engineers}
}

@article{patsialis2021multi,
  title={Multi-fidelity Monte Carlo for seismic risk assessment applications},
  author={Patsialis, D and Taflanidis, A},
  journal={Structural Safety},
  volume={93},
  pages={102129},
  year={2021},
  publisher={Elsevier}
}

@article{patsialis2020reduced,
  title={Reduced order modeling of hysteretic structural response and applications to seismic risk assessment},
  author={Patsialis, D and Taflanidis, A},
  journal={Engineering Structures},
  volume={209},
  pages={110135},
  year={2020},
  publisher={Elsevier}
}

@article{jalayer2020seismic,
  title={Seismic reliability assessment and the nonergodicity in the modelling parameter uncertainties},
  author={Jalayer, F  and Ebrahimian, H },
  journal={Earthquake Engineering \& Structural Dynamics},
  volume={49},
  number={5},
  pages={434--457},
  year={2020},
  publisher={Wiley Online Library}
}

@article{chen2010stochastic,
  title={Stochastic seismic response analysis of structures exhibiting high nonlinearity},
  author={Chen, J and Li, J},
  journal={Computers \& structures},
  volume={88},
  number={7-8},
  pages={395--412},
  year={2010},
  publisher={Elsevier}
}

@article{dos2016incremental,
  title={Incremental dynamic analysis: a nonlinear stochastic dynamics perspective},
  author={dos Santos, KRM and Kougioumtzoglou, IA and Beck, AT},
  journal={Journal of Engineering Mechanics},
  volume={142},
  number={10},
  pages={06016007},
  year={2016},
  publisher={American Society of Civil Engineers}
}

@article{gidaris2015kriging,
  title={Kriging metamodeling in seismic risk assessment based on stochastic ground motion models},
  author={Gidaris, I  and Taflanidis, A  and Mavroeidis, GP},
  journal={Earthquake Engineering \& Structural Dynamics},
  volume={44},
  number={14},
  pages={2377--2399},
  year={2015},
  publisher={Wiley Online Library}
}

@inproceedings{zou2020surrogate,
  title={Surrogate modeling for the seismic response estimation of residential wood frame structures},
  author={Zou, J and Welch, DP and Zsarnoczay, A and Taflanidis, A and Deierlein, G},
  booktitle={Proceedings of the 17th World Conference on Earthquake Engineering, Japan},
  year={2020}
}

@article{abbiati2021seismic,
  title={Seismic fragility analysis based on artificial ground motions and surrogate modeling of validated structural simulators},
  author={Abbiati, G  and Broccardo, M  and Abdallah, I  and Marelli, S and Paolacci, F},
  journal={Earthquake Engineering \& Structural Dynamics},
  volume={50},
  number={9},
  pages={2314--2333},
  year={2021},
  publisher={Wiley Online Library}
}

@article{kyprioti2021kriging,
  title={Kriging metamodeling for seismic response distribution estimation},
  author={Kyprioti, AP and Taflanidis, A },
  journal={Earthquake Engineering \& Structural Dynamics},
  volume={50},
  number={13},
  pages={3550--3576},
  year={2021},
  publisher={Wiley Online Library}
}

@article{giovanis2016epistemic,
  title={Epistemic uncertainty assessment using incremental dynamic analysis and neural networks},
  author={Giovanis, DG and Fragiadakis, M  and Papadopoulos, V },
  journal={Bulletin of Earthquake Engineering},
  volume={14},
  pages={529--547},
  year={2016},
  publisher={Springer}
}

@article{pang2014artificial,
  title={An artificial neural network based method for seismic fragility analysis of highway bridges},
  author={Pang, Y and Dang, X  and Yuan, W },
  journal={Advances in Structural Engineering},
  volume={17},
  number={3},
  pages={413--428},
  year={2014},
  publisher={SAGE Publications Sage UK: London, England}
}

@article{bakalis2017seismic,
  title={Seismic risk assessment of liquid storage tanks via a nonlinear surrogate model},
  author={Bakalis, K  and Vamvatsikos, D  and Fragiadakis, M},
  journal={Earthquake Engineering \& Structural Dynamics},
  volume={46},
  number={15},
  pages={2851--2868},
  year={2017},
  publisher={Wiley Online Library}
}

@article{zhu2023seismic,
  title={Seismic fragility analysis using stochastic polynomial chaos expansions},
  author={Zhu, X and Broccardo, M and Sudret, B },
  journal={Probabilistic Engineering Mechanics},
  pages={103413},
  year={2023},
  publisher={Elsevier}
}

@article{yi2024stochastic,
  title={Stochastic emulation with enhanced partial-and no-replication strategies for seismic response distribution estimation},
  author={Yi, S  and Taflanidis, A },
  journal={Earthquake Engineering \& Structural Dynamics},
  volume={53},
  number={7},
  pages={2354--2381},
  year={2024},
  publisher={Wiley Online Library}
}

@article{soize2016data,
  title={Data-driven probability concentration and sampling on manifold},
  author={Soize, C and Ghanem, R},
  journal={Journal of Computational Physics},
  volume={321},
  pages={242--258},
  year={2016},
  publisher={Elsevier}
}

@article{coifman2006geometric,
  title={Geometric harmonics: a novel tool for multiscale out-of-sample extension of empirical functions},
  author={Coifman, R and Lafon, S},
  journal={Applied and Computational Harmonic Analysis},
  volume={21},
  number={1},
  pages={31--52},
  year={2006},
}

@article{dsilva2018parsimonious,
  title={Parsimonious representation of nonlinear dynamical systems through manifold learning: A chemotaxis case study},
  author={Dsilva, Carmeline J and Talmon, Ronen and Coifman, Ronald R and Kevrekidis, Ioannis G},
  journal={Applied and Computational Harmonic Analysis},
  volume={44},
  number={3},
  pages={759--773},
  year={2018},
  publisher={Elsevier}
}

@article{iman1982distribution,
  title={A distribution-free approach to inducing rank correlation among input variables},
  author={Iman, Ronald L and Conover, William-Jay},
  journal={Communications in Statistics-Simulation and computation},
  volume={11},
  number={3},
  pages={311--334},
  year={1982},
  publisher={Taylor \& Francis}
}

@article{foutch2002modeling,
  title={Modeling of steel moment frames for seismic loads},
  author={Foutch, Douglas A and Yun, Seung-Yul},
  journal={Journal of Constructional Steel Research},
  volume={58},
  number={5-8},
  pages={529--564},
  year={2002},
  publisher={Elsevier}
}

@article{vamvatsikos2010incremental,
  title={Incremental dynamic analysis for estimating seismic performance sensitivity and uncertainty},
  author={Vamvatsikos, Dimitrios and Fragiadakis, Michalis},
  journal={Earthquake engineering \& structural dynamics},
  volume={39},
  number={2},
  pages={141--163},
  year={2010},
  publisher={Wiley Online Library}
}

@article{vamvatsikos2004applied,
  title={Applied incremental dynamic analysis},
  author={Vamvatsikos, Dimitrios and Cornell, C Allin},
  journal={Earthquake spectra},
  volume={20},
  number={2},
  pages={523--553},
  year={2004},
  publisher={SAGE Publications Sage UK: London, England}
}

@article{crabtree2025generative,
  title   = {Generative learning for slow manifolds and bifurcation diagrams},
  author  = {Crabtree, Ellis R. and Giovanis, Dimitris G. and Evangelou, Nikolaos and Bello-Rivas, Juan M. and Kevrekidis, Ioannis G.},
  journal = {Computers \& Chemical Engineering},
  volume  = {207},
  pages   = {109544},
  year    = {2026},
  doi     = {10.1016/j.compchemeng.2025.109544},
  publisher = {Elsevier}
}

\end{document}